\documentclass[pra, twocolumn,preprintnumbers,amsmath,amssymb,tightenlines,superscriptaddress]{revtex4}

\usepackage{graphicx}% Include figure files
\usepackage{dcolumn}% Align table columns on decimal point
\usepackage{bm}% bold math
\usepackage{epsfig}
\usepackage{color}
\usepackage{stmaryrd}
\usepackage{mathrsfs}
\usepackage[cal=zapfc]{mathalfa}
\usepackage{amsmath}
\usepackage{mdframed}
\usepackage{tcolorbox}
\usepackage[normalem]{ulem}
\usepackage{xcolor}

\newcommand{\bra}[1]{\langle{#1}|}
\newcommand{\ket}[1]{|{#1}\rangle}

\def\beq{\begin{equation}}
\def\eeq{\end{equation}}

\newcommand{\rref}[1]{Ref.~\cite{#1}}
\newcommand{\fref}[1]{Fig.~\ref{#1}}
\newcommand{\frefp}[2]{Fig.~\ref{#1}~(#2)}

\newcommand{\eref}[1]{Eq.~(\ref{#1})}
\newcommand{\esref}[2]{Eqs.~(\ref{#1}) and (\ref{#2})}

\newcommand{\sref}[1]{section~\ref{#1}}

\newcommand{\cref}[1]{chapter~\ref{#1}}

\newcommand{\Cref}[1]{Chapter~\ref{#1}}

\newcommand{\aref}[1]{appendix~\ref{#1}}

\usepackage{float}

\begin{document}

\title{Adiabatic controlled-$Z$ gate for neutral atoms with Förster-enhanced Rydberg dressing}
\author{Diksha Thapliyal }
\affiliation{Department of Physics}

\author{Ajay Wasan}
\email{awasan@ph.iitr.ac.in}
\affiliation{Department of Physics}
\affiliation{Centre for Photonics and Quantum Communication Technology, Indian Institute of Technology Roorkee, 247667, India}

\begin{abstract}
We study an adiabatic controlled-Z (CZ) gate for neutral atoms in which dipole-dipole coupling between nearly degenerate Rydberg pair states enhances the interaction between dressed ground states. This coupling produces an avoided crossing in the two-atom dressed-state spectrum. Approaching the crossing enhances the dressed interaction, but also reduces the dressed-state gap and increases the Rydberg admixture. We identify a F\"orster-enhanced operating range in which the required conditional phase is accumulated with adiabatic return and small transient Rydberg population. We then analyze the finite-time dynamics of the two-qubit computational states under a smooth pulse. Reducing the dressing strength weakens the dressed interaction, so maintaining the target phase at the same pulse duration requires operation closer to the avoided crossing, where the Rydberg admixture increases. This dependence is then used to select the shortest gate at a fixed peak Rabi frequency in the F\"orster-enhanced operating range that accumulates the CZ phase with high Bell-state fidelity while keeping the Rydberg population small. Changing the F\"orster defect shifts the avoided crossing, while the gate parameters change only weakly. For a specific rubidium F\"orster channel, the selected gate corresponds to experimentally relevant laser coupling, pulse duration, and interatomic separation. We also estimate errors from the finite Rydberg lifetime and from weak excitation of a neighboring spectator atom. For the parameters considered, the spectator-induced error remains below the finite-lifetime contribution.
\end{abstract}

\maketitle

%%%%%%%%%%%%%%%%%%%%%%%%%%%%%%%%%%%%%%%%%%%%%%%%%%%
\section{Introduction}
%%%%%%%%%%%%%%%%%%%%%%%%%%%%%%%%%%%%%%%%%%%%%%%%%%%
\label{sec:introduction}

Optical tweezers trap neutral atoms individually and allow their rearrangement into nearly defect-free arrays with programmable geometries in one, two, and three dimensions \cite{endres2016atom,barredo2016atom,barredo2018synthetic}. In these arrays, long-lived ground states encode the qubits, while optical and microwave fields provide state preparation and coherent control \cite{jaksch2000fast,saffman2010quantum,browaeys2020many,henriet2020quantum, morgado2021quantum,shi2022quantum}. This has enabled programmable quantum simulation and quantum information processing with trapped neutral atoms \cite{bernien2017probing,ebadi2021quantum,bluvstein2022quantum, bluvstein2024logical}. Entangling gates require interactions between selected qubits. Excitation to Rydberg states provides these interactions, as proposed early in the development of neutral atom quantum computing \cite{jaksch2000fast} and subsequently demonstrated in experiments on Rydberg blockade and two-qubit gates \cite{gaetan2009observation,isenhower2010demonstration,maller2015rydberg,levine2018high}.

Several approaches have been developed to implement Rydberg-mediated two-qubit gates. These include resonant blockade sequences, adiabatic protocols, and gates based on optimized pulses \cite{levine2019parallel,saffman2020symmetric,petrosyan2017high, sun2020controlled,sun2023off,jandura2022time,graham2019rydberg, fu2022high,evered2023high}. These approaches differ in their transient Rydberg state population and in how the Rydberg interaction generates the conditional phase. Since population in the Rydberg states can be lost through spontaneous decay, the finite Rydberg lifetime remains a significant source of gate error.

In adiabatic Rydberg dressing, entangling gates can be implemented without complete population transfer to a bare Rydberg state \cite{keating2015robust,mitra2020robust,mitra2023neutral}. A smooth laser pulse admixes a Rydberg component into one of the ground-state qubit levels, and the system adiabatically follows the resulting dressed state. When both atoms are dressed, the interaction between their Rydberg components shifts the energy of the two-atom dressed state relative to that expected from independent single-atom dressing \cite{johnson2010interactions,keating2013adiabatic}. This interaction-induced energy shift produces an additional phase on the two-atom state, providing the conditional phase required for the gate.

The small Rydberg admixture is an important advantage of the dressing approach because it reduces the Rydberg state population during the pulse and hence the effect of spontaneous decay. If the evolution remains adiabatic throughout the pulse, the atoms return to the computational states without residual Rydberg excitation at the end of the pulse \cite{keating2015robust,mitra2023neutral}. The same small admixture, however, weakens the dressed interaction, so a longer pulse is required to accumulate the conditional phase. The finite-lifetime error therefore depends not only on the Rydberg population during the pulse but also on the gate duration.

A nearby Rydberg pair state can modify the balance between the dressed interaction and the Rydberg state population. When the doubly excited Rydberg state $\ket{rr}$ lies close in energy to another Rydberg pair state $\ket{pp'}$, the dipole-dipole interaction couples the two states, while the F\"orster defect sets their bare energy separation \cite{walker2008consequences,ravets2015measurement}. F\"orster-coupled pair dynamics have been proposed for gates based on direct Rydberg excitation \cite{beterov2016adiabatic,huang2018robust}. Near-degenerate pair channels have been shown to affect population leakage and gate fidelity in finite blockade gates \cite{thapliyal2025controlled}. Explicit treatment of the coupled pair states has also been shown to change predicted gate fidelities across different protocols \cite{norrell2026entangling}. In Rydberg dressed systems, F\"orster coupling can strongly enhance the effective interaction between the dressed ground states \cite{vanbijnen2015quantum}.

For an adiabatic dressing gate, however, an enhanced dressed interaction alone does not determine the gate performance. Previous analyses beyond the perfect-blockade regime have represented the Rydberg-Rydberg interaction as a finite energy shift of the doubly excited state $\ket{rr}$ \cite{mitra2023neutral}. Near a F\"orster resonance, the dipole-dipole interaction couples $\ket{rr}$ to the near-degenerate pair state $\ket{pp'}$, so the gate dynamics depend on the coupled pair states rather than on an effective shift of $\ket{rr}$ alone. The same changes in the pair-state spectrum that enhance the interaction also affect the Rydberg admixture and the energy gap relevant to adiabatic following. For a fixed pulse duration, the dressing strength and the proximity to the avoided crossing must be selected together to maintain the required conditional phase. 

In this work, we study a F\"orster-enhanced adiabatic controlled-$Z$ (CZ) gate driven by a smooth pulse with the laser detuning held fixed during each pulse. We retain the dipole-dipole coupling between $\ket{rr}$ and the near-degenerate pair state $\ket{pp'}$ to examine both the dressed interaction and the Rydberg population during the pulse. We identify a F\"orster-enhanced operating range in which the dressed interaction is increased while adiabatic return is maintained. Reducing the dressing strength suppresses the laser-induced Rydberg admixture but also weakens the interaction. Accumulating the target conditional phase within a fixed pulse duration can therefore require operation near the avoided crossing, where the pair-state admixture increases. Within the pulse and parameter ranges considered, we select the shortest gate at fixed peak Rabi frequency that accumulates the CZ phase with adiabatic return and high Bell-state fidelity while keeping the peak Rydberg population below a chosen upper limit. We then examine how the operating conditions change with the F\"orster defect and relate the selected gates to a specific rubidium F\"orster channel. Finally, we estimate errors from finite Rydberg lifetimes and weak illumination of a neighboring spectator atom, with spectator interactions represented by diagonal energy shifts.

The paper is organized as follows. In \sref{sec:dressed_mechanism}, we introduce the two-atom model and analyze the dressed state mechanism that determines the F\"orster-enhanced operating range. In \sref{sec:cz_dynamics}, we study the finite-time CZ-gate dynamics, determine the gate operating point, and analyze its dependence on the F\"orster defect. In \sref{sec:decay}, we include the error due to finite Rydberg lifetimes, and in \sref{sec:crosstalk}, we analyze the effect of a weakly driven spectator atom on the gate. Additional details of the equations of motion, representative gate dynamics, and the three-atom spectator model are given in the Appendices.

%%%%%%%%%%%%%%%%%%%%%%%%%%%%%%%%%%%%%%%%%%%%%%%%%%
\section{Rydberg dressed interaction and operating range}
\label{sec:dressed_mechanism}
%%%%%%%%%%%%%%%%%%%%%%%%%%%%%%%%%%%%%%%%%%%%%%%%%%%

%%%%%%%%%%%%%%%%%%%%%%%%%%%%%%%%%%%%%%%%%%%%%%%%%%%
\subsection{Level structure and Hamiltonian}
%%%%%%%%%%%%%%%%%%%%%%%%%%%%%%%%%%%%%%%%%%%%%%%%%%%

\begin{figure*}[htb]
\centering
\includegraphics[width=\textwidth]{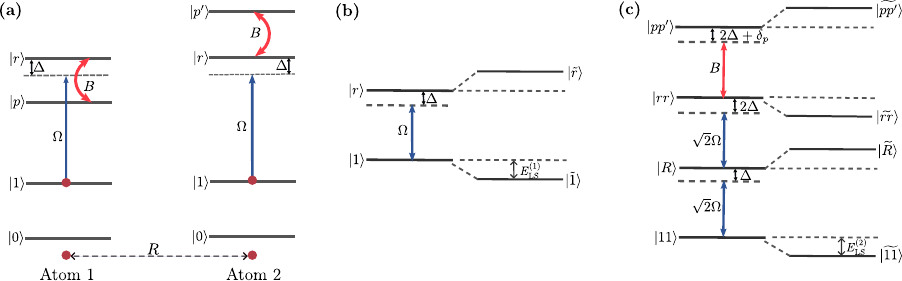}
\caption{
Energy-level structure and Rydberg dressing for the CZ gate. (a) Bare level structure of two atoms separated by the interatomic distance $R$. The laser couples $\ket{1}$ to $\ket{r}$ with Rabi frequency $\Omega$ and detuning $\Delta$, while the dipole-dipole interaction couples the pair states $\ket{rr}$ and $\ket{pp'}$ with strength $B$. (b) Single-atom dressing in the basis $\{\ket{1},\ket{r}\}$. The dressed state $\ket{\widetilde{1}}$ connected to $\ket{1}$ acquires the light shift $E_{\rm LS}^{(1)}$. (c) Two-atom dressing in the basis $\{\ket{11},\ket{R},\ket{rr},\ket{pp'}\}$, where $\ket{R}=(\ket{1r}+\ket{r1})/\sqrt{2}$. The dressed state $\ket{\widetilde{11}}$ connected to $\ket{11}$ acquires the two-atom light shift $E_{\rm LS}^{(2)}$.
}
\label{fig:model}
\end{figure*}

The energy-level structure used to implement the two-qubit controlled-$Z$ gate is shown in \frefp{fig:model}{a}. The qubit states $\ket{0}$ and $\ket{1}$ are encoded in long-lived hyperfine ground states of two alkali atoms. The state $\ket{1}$ is coupled off-resonantly to a Rydberg state $\ket{r}$, whereas $\ket{0}$ remains uncoupled from the driving field. Consequently, the computational basis state $\ket{00}$ is unaffected by the drive. For the states $\ket{01}$ and $\ket{10}$, only the atom in $\ket{1}$ is driven, and the dynamics reduce to the two-level system shown in \frefp{fig:model}{b}. In the rotating frame, the corresponding single-atom Hamiltonian is
\begin{equation}
\frac{H_{1}(t)}{\hbar}
=
\frac{\Omega(t)}{2}
\big[
\ket{1}\bra{r}
+
\ket{r}\bra{1}
\big]
+
\Delta\ket{r}\bra{r},
\label{eq:single_atom_hamiltonian}
\end{equation}
in the basis $\{\ket{1},\ket{r}\}$, where $\Omega(t)$ is the time-dependent single-atom Rabi frequency and $\Delta$ is the laser detuning.
For the initial state $\ket{11}$, symmetric driving couples $\ket{11}$ to the singly excited state $\ket{R} = (\ket{1r} + \ket{r1})/\sqrt{2}$. The state $\ket{R}$ couples to the doubly excited state $\ket{rr}$, which is in turn coupled by the resonant dipole-dipole interaction strength $B$ to a near-degenerate Rydberg pair state $\ket{pp'}$ that is not directly driven by the laser. The resulting four-level linkage is shown in \frefp{fig:model}{c}. In the basis $\{\ket{11},\ket{R},\ket{rr},\ket{pp'}\}$, the two-atom Hamiltonian is
\begin{align}
\frac{H_{11}(t)}{\hbar}
={}&
\frac{\Omega(t)}{\sqrt{2}}
\big[ \ket{11}\bra{R} + \ket{R}\bra{rr} + \mathrm{H.c.} \big]
\nonumber\\
&+B \big[ \ket{rr}\bra{pp'} + \mathrm{H.c.} \big]
\nonumber\\
&+ \Delta\ket{R}\bra{R} + 2\Delta\ket{rr}\bra{rr}
\nonumber\\
&+ (2\Delta+\delta_p) \ket{pp'}\bra{pp'}
\label{eq:two_atom_hamiltonian}
\end{align}
The F\"orster defect $\delta_p$ is defined as $\delta_p = (E_p+E_{p'}-2E_r)/\hbar$ where $E_r$, $E_p$, and $E_{p'}$ are the bare energies of the corresponding Rydberg states. We retain only the near-resonant Rydberg pair channel $\ket{rr}\leftrightarrow\ket{pp'}$ and neglect couplings to other far-detuned Rydberg pair states. For fixed pair states and relative atomic orientation, the dipole-dipole coupling $B$ scales as $R^{-3}$, where $R$ is the interatomic separation.

%%%%%%%%%%%%%%%%%%%%%%%%%%%%%%%%%%%%%%%%%%%%%%%%
\subsection{Rydberg dressed interaction}
%%%%%%%%%%%%%%%%%%%%%%%%%%%%%%%%%%%%%%%%%%%%%%%%

During an adiabatic pulse, the single and two-qubit computational input states follow instantaneous eigenstates of the Hamiltonians in \eref{eq:single_atom_hamiltonian} and \eref{eq:two_atom_hamiltonian}, respectively. We denote by $\ket{\tilde{1}(t)}$ the eigenstate of $H_1(t)$ that approaches $\ket{1}$ as $\Omega(t)\rightarrow 0$, and by $\ket{\widetilde{11}(t)}$ the eigenstate of $H_{11}(t)$ that approaches $\ket{11}$ in the same limit. Their corresponding eigenenergies are $E_{\tilde{1}}(t)$ and $E_{\widetilde{11}}(t)$. The associated single and two-atom light shifts measured relative to the bare computational states energies $E_{1}$ and $E_{11}$, are $E_{\rm LS}^{(1)}(t) = E_{\widetilde{1}}(t)-E_{1}$ and $E_{\rm LS}^{(2)}(t) = E_{\widetilde{11}}(t)-E_{11}$, respectively, as illustrated in \fref{fig:model}.

In the absence of interatomic coupling, the atoms are dressed independently, so the two-atom light shift satisfies $E_{\rm LS}^{(2)}(t)=2E_{\rm LS}^{(1)}(t)$. We therefore define the interaction-induced contribution to the two-atom light shift as the Rydberg dressed interaction energy,
\begin{equation}
\hbar\kappa(t)
=E_{\rm LS}^{(2)}(t)-2E_{\rm LS}^{(1)}(t).
\label{eq:kappa}
\end{equation}
In the present model, $B=0$ corresponds to independent dressing and hence $\kappa(t)=0$. For nonzero $B$, the dipole-dipole coupling between $\ket{rr}$ and the near-degenerate pair state $\ket{pp'}$ shifts the energy of $\ket{\widetilde{11} (t)}$, while the single-atom light shift remains unchanged, resulting in a nonzero $\kappa(t)$.

\begin{figure*}[htb]
\centering
\includegraphics[width=0.98\textwidth]{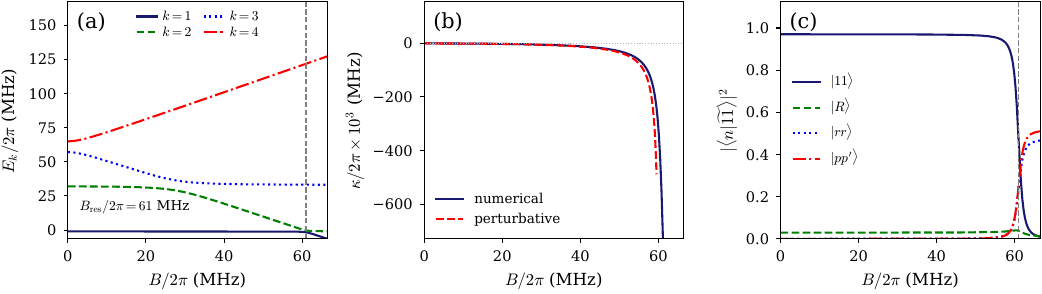}
\caption{
Dressed state spectrum as a function of the pair coupling $B$, for $\Omega/2\pi=8~{\rm MHz}$, $\Delta/2\pi=32~{\rm MHz}$, and $\delta_p/2\pi=-6.9~{\rm MHz}$. (a) Dressed eigenenergies $E_k/2\pi$ as functions of $B/2\pi$. The vertical dashed line indicates $B_{\rm res}/2\pi=61~{\rm MHz}$, defined by the minimum energy gap between $\ket{\widetilde{11}}$ and the nearest dressed state. (b) Effective dressed interaction $\kappa$ obtained from the full Hamiltonian (blue solid) and from the perturbative expression in \eref{eq:kappa_pert} (red dashed). (c) Bare-state probabilities of $\ket{\widetilde{11}}$ for the basis states $\ket{11}$, $\ket{R}$, $\ket{rr}$, and $\ket{pp'}$.
}
\label{fig:mechanism}
\end{figure*}

For a time-dependent pulse, the light shifts $E_{\rm LS}^{(1)}(t)$ and $E_{\rm LS}^{(2)}(t)$, and hence $\kappa(t)$, vary with time. Under adiabatic evolution, the conditional phase accumulated over a pulse duration $T$ is
\begin{equation}
    \phi(T) = - \int_0^T \kappa(t) \, dt,
    \label{eq:conditional_phase}
\end{equation} 
which is the phase acquired by $\ket{11}$ relative to the sum of the phases acquired by the single excitation states $\ket{01}$ and $\ket{10}$.

If the evolution returns each computational basis state to itself up to a phase, the single-qubit phases can be removed by local $Z$ rotations. The remaining two-qubit operation is then determined by the conditional phase $\phi(T)$. For $\phi(T) = \pi$ (mod $2\pi$), the operation is locally equivalent to a controlled-$Z$ gate. In the computational basis the ideal CZ gate is 
\begin{equation}
    U_{\rm CZ}=\operatorname{diag}(1,1,1,-1).
\end{equation}

Equation~\eqref{eq:conditional_phase} shows that a larger $|\kappa|$ can reduce the time required to reach the target conditional phase. To determine how the F\"orster coupling changes $\kappa$, we first consider a fixed $\Omega$ in the weak-dressing regime, $\Omega/\Delta\ll1$. In this limit, the dressed energies can be expanded perturbatively in powers of $\Omega$. The leading nonzero contribution to $\kappa$ is fourth order in the laser coupling and is given by
\begin{equation}
\kappa \approx -\frac{\Omega^4}{4\Delta^2}
\left[
\frac{1}{2\Delta-\dfrac{B^2}{2\Delta+\delta_p}}
-
\frac{1}{2\Delta}
\right]
\label{eq:kappa_pert}
\end{equation}
where the first term gives the two-atom contribution in the presence of the F\"orster coupling, while the second term gives the corresponding contribution for independent dressing ($B=0$). The F\"orster coupling enters through the denominator $2\Delta-B^2/(2\Delta+\delta_p)$. 

For fixed $\Omega$ and $\Delta$, the effective interaction $|\kappa|$ increases with  $B$, and at a given pulse duration, the required conditional phase can be reached with a smaller laser coupling. However, when $B=B_{\rm res}^{(0)}=\sqrt{2\Delta(2\Delta+\delta_p)}$, the denominator in \eref{eq:kappa_pert} vanishes, and the perturbative result is no longer valid. Hence, we next use the full two-atom Hamiltonian to analyze the dressed state spectrum and interaction as $B$ is varied.

%%%%%%%%%%%%%%%%%%%%%%%%%%%%%%%%%%%%%%%%%%%%%%%%
\subsection{Dressed state spectrum and operating range}
\label{sec:spectrum}
%%%%%%%%%%%%%%%%%%%%%%%%%%%%%%%%%%%%%%%%%%%%%%%%

For gate operation, the magnitude of $\kappa$ alone is not sufficient to determine the relevant range of $B$. The energy gap between $\ket{\widetilde{11}(t)}$ and the nearest dressed state is important for adiabatic following, while the Rydberg admixture of $\ket{\widetilde{11}(t)}$ gives the Rydberg component carried by the dressed state. We therefore consider these quantities together with $\kappa$ as the pair coupling is varied. To analyze their dependence on $B$, we first take the laser coupling $\Omega$ to be fixed, so that the Hamiltonian is time independent. The dressed states $\ket{\varphi_k}$ and their eigenenergies $E_k$ are obtained from
\begin{equation}
H_{11}\ket{\varphi_k}
=
E_k\ket{\varphi_k},
\qquad
k=1,\ldots,4 .
\label{eq:dressed_eigenproblem}
\end{equation}

Each $\ket{\varphi_k}$ is a superposition of the bare states $\{\ket{11},\ket{R},\ket{rr},\ket{pp'}\}$. Among these dressed states, we are interested in the state $\ket{\widetilde{11}}\,(k=1)$ and its separation from the neighboring dressed states. We therefore define the minimum energy gap as
\begin{equation}
E_{\rm gap}(B)
=
\min_{k\neq1}
\left|E_k(B)-E_{\widetilde{11}}(B)\right|.
\label{eq:gap}
\end{equation}
To illustrate the dependence on the pair coupling, we take
$\delta_p/2\pi=-6.9~{\rm MHz}$ for the $^{87}\mathrm{Rb}$ pair channel and fix $\Omega/2\pi=8~{\rm MHz}$ and $\Delta/2\pi=32~{\rm MHz}$. 

\begin{figure}[htb]
\centering
\includegraphics[width=\columnwidth]{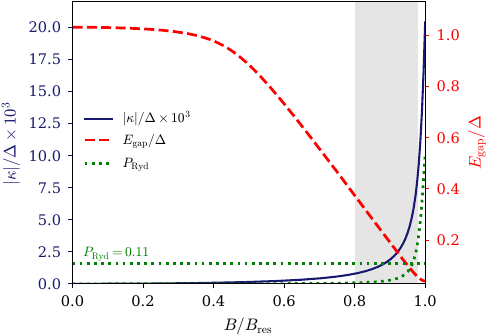}
\caption{
Effective dressed interaction $|\kappa|/\Delta\times10^{3}$ (blue solid, left axis), dressed state gap $E_{\rm gap}/\Delta$ between $\ket{\widetilde{11}}$ and the nearest dressed state (red dashed, right axis), and Rydberg admixture $P_{\rm Ryd}$ of $\ket{\widetilde{11}}$ (green dotted, right axis), as functions of the normalized pair coupling $B/B_{\rm res}$ for the same parameters as in \fref{fig:mechanism}. Here, $B_{\rm res}$ is defined by the minimum of $E_{\rm gap}$. The horizontal dotted line shows reference value $P_{\rm Ryd}=0.11$, and the shaded region indicates the interval $B/B_{\rm res}\in[0.80,0.98]$.
}
\label{fig:tradeoff}
\end{figure}

The dressed state energies are shown in \frefp{fig:mechanism}{a}, as a function of $B$. At small $B$, $E_{\widetilde{11}}\,(k=1)$ is well separated from the other dressed states. As $B$ increases, the energy separation between $\ket{\widetilde{11}}$ and a neighboring dressed state decreases, and for the laser coupling considered here the two states form an avoided crossing. We denote the corresponding pair coupling by $B_{\rm res}$, defined as the value of $B$ at which $E_{\rm gap}$ is minimum. For the parameter set considered here, $B_{\rm res}/2\pi\simeq61~{\rm MHz}$. For a given pulse variation, the smaller gap enhances the nonadiabatic coupling between $\ket{\widetilde{11}}$ and the neighboring dressed state, increasing the probability of transition between them. 

In \frefp{fig:mechanism}{b}, we show the interaction $\kappa$ obtained from the full Hamiltonian together with the perturbative result given in \eref{eq:kappa_pert}. The two results agree away from the avoided crossing, while the perturbative result deviates near $B_{\rm res}^{(0)}$. The numerical value of $|\kappa|$ remains finite and increases toward the avoided crossing.

The bare state composition of $\ket{\widetilde{11}}$ is shown in \frefp{fig:mechanism}{c}. At small $B$, $\ket{\widetilde{11}}$ is predominantly $\ket{11}$, with only a small admixture of the Rydberg-containing states due to the finite laser coupling. Closer to $B_{\rm res}$, the contribution from $\ket{11}$ decreases, while those from $\ket{rr}$ and $\ket{pp'}$ increase. Thus, the enhancement of $|\kappa|$ near the avoided crossing is accompanied by a larger Rydberg-pair admixture in $\ket{\widetilde{11}}$.

To quantify the Rydberg admixture in $\ket{\widetilde{11}}$, we define
\begin{equation}
P_{\rm Ryd}
=
|\langle R|\widetilde{11}\rangle|^2
+
|\langle rr|\widetilde{11}\rangle|^2
+
|\langle pp'|\widetilde{11}\rangle|^2 ,
\label{eq:Pryd_unweighted}
\end{equation}
which gives the total probability of the Rydberg components in $\ket{\widetilde{11}}$.

In \fref{fig:tradeoff}, we show $|\kappa|$, $E_{\rm gap}$, and $P_{\rm Ryd}$ as functions of $B/B_{\rm res}$. As $B$ approaches $B_{\rm res}$, the dressed interaction $|\kappa|$ increases, but $E_{\rm gap}$ decreases rapidly and $P_{\rm Ryd}$ grows. Operating closer to the avoided crossing would therefore increase the Rydberg pair admixture and, for a given pulse variation, increase the probability of nonadiabatic transitions to the neighboring dressed state. We refer to the shaded interval as the F\"orster-enhanced operating range, where the F\"orster coupling enhances the dressed interaction before the rapid decrease of the energy gap and increase of the Rydberg admixture near the avoided crossing. Hence, we restrict the gate analysis to this range on the lower pair coupling side of the avoided crossing. Within this range, the gate parameters are determined from the time-dependent dynamics by requiring the target conditional phase together with population return and adiabatic evolution.

%%%%%%%%%%%%%%%%%%%%%%%%%%%%%%%%%%%%%%%%%%%%%%%%
\section{Time-dependent CZ dynamics}
\label{sec:cz_dynamics}
%%%%%%%%%%%%%%%%%%%%%%%%%%%%%%%%%%%%%%%%%%%%%%%%

%%%%%%%%%%%%%%%%%%%%%%%%%%%%%%%%%%%%%%%%%%%%%%%%
\subsection{Pulse protocol and gate metrics}
\label{sec:pulse_metrics}
%%%%%%%%%%%%%%%%%%%%%%%%%%%%%%%%%%%%%%%%%%%%%%%%

The analysis in \sref{sec:spectrum} was carried out at fixed $\Omega$, $\Delta$, and $\delta_p$ to identify the F\"orster-enhanced operating range from the dressed state spectrum. We now study the finite duration CZ dynamics in this range. For each pulse, $\Delta$, $B$, and $\delta_p$ are held fixed, while the Rabi frequency is varied as
\begin{equation}
\Omega(t)
=
\Omega_{\max}\sin^2\!\left(\frac{\pi t}{T}\right),
\qquad 0\leq t\leq T ,
\label{eq:pulse}
\end{equation}
where $T$ is the pulse duration. The time evolution of the single and two-atom states is obtained by solving the Schr\"odinger equations for $H_1(t)$ and $H_{11}(t)$, starting from $\ket{\psi_1(0)}=\ket{1}$ and $\ket{\psi_{11}(0)}=\ket{11}$, as described in \aref{app:eom_1}.

Since $\Omega(0)=\Omega(T)=0$, the dressed states $\ket{\widetilde{1}(t)}$ and $\ket{\widetilde{11}(t)}$ approach $\ket{1}$ and $\ket{11}$, respectively, at the beginning and end of the pulse. Under adiabatic evolution, the system follows these instantaneous dressed states throughout the pulse and returns to the initial computational states with accumulated phases. The $\sin^2$ envelope turns the laser coupling on and off smoothly, with $\dot{\Omega}(0)=\dot{\Omega}(T)=0$, which suppresses rapid changes at the pulse boundaries.

At the end of the pulse, the input state $\ket{00}$ is unchanged. For the input states $\ket{01}$ and $\ket{10}$, the single-atom return amplitude is $c_1=\langle1|\psi_1(T)\rangle$. For the input state $\ket{11}$, evolution under the four-level Hamiltonian gives the return amplitude $c_{11}=\langle11|\psi_{11}(T)\rangle$. The conditional phase obtained from these final amplitudes is
\begin{equation}
\phi_{\rm CZ}(T)
=
\arg(c_{11})-2\arg(c_1)
\quad (\mathrm{mod}\ 2\pi).
\label{eq:phi_cz}
\end{equation}
In the adiabatic limit, the above equation reduces to \eref{eq:conditional_phase}. A CZ gate, up to local $Z$ rotations, requires $\phi_{\rm CZ} (T)=\pi\;(\mathrm{mod}\;2\pi)$ together with complete return to the computational basis states, $|c_1|^2=|c_{11}|^2=1$. For a finite duration pulse, deviations from these conditions give a conditional phase error and incomplete population return.

To quantify these effects together, we consider the input state $\ket{++}=(\ket{00}+\ket{01}+\ket{10}+\ket{11})/2$. At the end of the pulse, the states $\ket{01}$ and $\ket{10}$ have the same single-qubit phase $\theta_s=\arg(c_1)$, which we remove by applying $Z(\theta_s)=\operatorname{diag}(1,e^{-i\theta_s})$ to each qubit, so that the phase remaining on the $\ket{11}$ component is the conditional phase. The corrected final state in the computational basis is
\begin{equation}
\ket{\psi_{\rm corr}(T)} = \frac{1}{2}\left[
\ket{00}
+|c_1|\ket{01}
+|c_1|\ket{10}
+|c_{11}|e^{i\phi_{\rm CZ}}\ket{11}
\right].
\end{equation}
The ideal CZ output for the same input state is $\ket{\Phi_{\rm CZ}} =(\ket{00}+\ket{01}+\ket{10}-\ket{11})/2$. This state is locally equivalent to a Bell state up to a single-qubit rotation. The corresponding Bell state fidelity is,
\begin{equation}
F_{\rm Bell}
=
\left|
\langle\Phi_{\rm CZ}|\psi_{\rm corr}(T)\rangle
\right|^2.
\label{eq:Fbell_def}
\end{equation}
Although $F_{\mathrm{Bell}}$ measures the final-state fidelity for the input $\ket{++}$, it does not determine whether the evolution remains adiabatic during the pulse. The states $\ket{01}$ and $\ket{10}$ undergo only the single-atom dressing dynamics, whereas the $\ket{11}$ state follows the interacting two-atom dynamics and is affected by the F\"orster-induced avoided crossing. We therefore examine adiabatic following for the evolution starting from $\ket{11}$, which should follow the instantaneous dressed state $\ket{\widetilde{11}(t)}$ throughout the pulse. We quantify this by
\begin{equation}
P_{\rm adiab}^{\min}
=
\min_{0\leq t\leq T}
\left|
\langle \widetilde{11}(t)|\psi_{11}(t)\rangle
\right|^2 ,
\label{eq:Padiab}
\end{equation}
where $\ket{\psi_{11}(t)}$ denotes the two-atom state at time $t$.  A value $P_{\rm adiab}^{\min}=1$ corresponds to exact adiabatic following throughout the pulse, while smaller values indicate nonadiabatic transitions from $\ket{\widetilde{11}(t)}$ to other instantaneous dressed states. Such transitions can change the accumulated phase and lead to incomplete return to $\ket{11}$ at the end of the pulse.

Adiabatic following alone is not sufficient to characterize the gate dynamics, because the instantaneous dressed state $\ket{\widetilde{11}(t)}$ itself contains Rydberg admixture. Since the Rydberg states have finite lifetimes, their transient population during the pulse can contribute to the gate error. Hence, we also calculate the Rydberg state population of the evolving state,
\begin{equation}
P_{\rm Ryd}(t)
=
|\langle R|\psi_{11}(t)\rangle|^2
+
|\langle rr|\psi_{11}(t)\rangle|^2
+
|\langle pp'|\psi_{11}(t)\rangle|^2,
\label{eq:PRyd_of_t}
\end{equation}
and record its maximum during the pulse,
\begin{equation}
P_{\rm Ryd}^{\max}
=
\max_{0\le t\le T} P_{\rm Ryd}(t).
\label{eq:PRydmax}
\end{equation}

Equations~\eqref{eq:Padiab} and \eqref{eq:PRydmax} characterize distinct aspects of the dynamics. A pulse may follow the instantaneous dressed state almost perfectly while still having a large $P_{\rm Ryd}^{\max}$, because the dressed state itself contains Rydberg admixture. Conversely, nonadiabatic transitions may occur even when the Rydberg population remains moderate. Hence, $P_{\rm adiab}^{\min}$ quantifies adiabatic following, whereas $P_{\rm Ryd}^{\max}$ measures the maximum Rydberg state population reached during the pulse.

\begin{figure*}[htb]
\centering
\includegraphics[width=\textwidth]{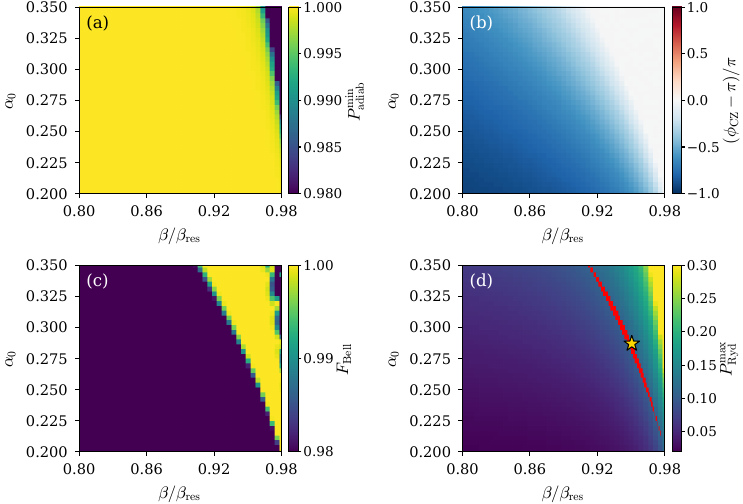}
\caption{
CZ-gate performance at zero relative F\"orster defect ($\gamma=0$) as functions of the peak dressing ratio $\alpha_0=\Omega_{\max}/\Delta$ and the normalized pair coupling $\beta/\beta_{\rm res}$. (a) Minimum adiabatic population $P_{\rm adiab}^{\min}$. (b) Conditional phase error $(\phi_{\rm CZ}-\pi)/\pi$. (c) Bell-state fidelity $F_{\rm Bell}$. (d) Peak Rydberg state population $P_{\rm Ryd}^{\max}$. For each $(\alpha_0,\beta/\beta_{\rm res})$, the quantities are evaluated at the smallest complete-pulse duration for which $\phi_{\rm CZ}(\tau_{\rm gate})=\pi$ within $\alpha_0\tau_{\rm gate}\leq160\pi$. The red region in (d) satisfies $\phi_{\rm CZ}=\pi$, $P_{\rm adiab}^{\min}\geq0.999$, $F_{\rm Bell}\geq0.9999$, and $P_{\rm Ryd}^{\max}\leq0.11$. The star marks the point with the smallest $\alpha_0\tau_{\rm gate}$ within this region.
}
\label{fig:gate_map}
\end{figure*}
%
%%%%%%%%%%%%%%%%%%%%%%%%%%%%%%%%%%%%%%%%%%%%%%%%%%%%%%%%%%%%%%%%%%
\subsection{Dimensionless parameters and physical constraints}
\label{sec:dimensionless}
%%%%%%%%%%%%%%%%%%%%%%%%%%%%%%%%%%%%%%%%%%%%%%%%%%%%%%%%%%%%%%%%%%

We scale all parameters by $\Delta$ and introduce the dimensionless time $\tau=\Delta t$. For the pulse in Eq.~\eqref{eq:pulse}, the coherent dynamics depends on the following dimensionless parameters
\begin{equation}
\alpha_0=\frac{\Omega_{\max}}{\Delta},
\qquad
\beta=\frac{B}{\Delta},
\qquad
\gamma=\frac{\delta_p}{\Delta},
\qquad
\tau_{\rm gate}=\Delta T .
\label{eq:dimensionless_params}
\end{equation}
Here, $\alpha_0$ is the peak dressing ratio and measures the laser coupling relative to the detuning. A larger $\alpha_0$ produces stronger Rydberg admixture at the pulse maximum. In the weak dressing regime, the single-atom Rydberg weight scales as $(\alpha_0/2)^2$, while \eref{eq:kappa_pert} gives $|\kappa|/\Delta\propto\alpha_0^4$. The normalized coupling $\beta/\beta_{\rm res}$ measures the proximity to the avoided crossing, with $\beta_{\rm res}=B_{\rm res}/\Delta$. For each $\alpha_0$ and $\gamma$, $\beta_{\rm res}$ is determined from the minimum of the dressed state gap. The relative defect $\gamma$ changes the separation between the pair states and therefore shifts the value of $B$ at which the avoided crossing occurs. The dimensionless duration $\tau_{\rm gate}$ sets both the time available to accumulate phase and how rapidly the pulse is varied. Increasing $\tau_{\rm gate}$ improves adiabatic following, but also increases the time spent with finite Rydberg admixture.

\begin{equation}
\tau_{\rm gate}
=
\frac{\Omega_{\max} T}{\alpha_0}.
\label{eq:physical_conversion}
\end{equation}
For analyzing the CZ-gate performance, we use $\tau_{\rm gate} \leq 160\pi/\alpha_0$, corresponding to $\Omega_{\max}/2\pi \leq 8~{\rm MHz}$ and $T \leq 10~\mu{\rm s}$. The chosen Rabi-frequency scale is consistent with experimentally demonstrated Rydberg excitation in $^{87}{\rm Rb}$ \cite{gaetan2009observation,wilk2010entanglement}. These values set the scale used for the gate analysis rather than a fundamental constraint of the Hamiltonian.

%%%%%%%%%%%%%%%%%%%%%%%%%%%%%%%%%%%%%%%%%%%%%%%%%%%%%%%%%%%%%%%%%%
\subsection{Gate performance at fixed F\"orster defect}
\label{sec:gate_map}
%%%%%%%%%%%%%%%%%%%%%%%%%%%%%%%%%%%%%%%%%%%%%%%%%%%%%%%%%%%%%%%%%%

We begin with the F\"orster-resonant case, $\gamma=0$, for which the bare Rydberg pair states $\ket{rr}$ and $\ket{pp'}$ are degenerate. The peak dressing ratio $\alpha_0$ is varied over $0.20$-$0.35$ and the normalized pair coupling $\beta/\beta_{\rm res}$ from $0.80$ to $0.98$. Since $\beta/\beta_{\rm res}=1$ corresponds to the position of the avoided crossing, this interval approaches the crossing from lower $\beta$ and lies within the F\"orster-enhanced operating range.

For each $(\alpha_0,\beta/\beta_{\rm res})$, we calculate the time evolution of the single and two-atom states. The conditional phase, final return probabilities, Bell-state fidelity, minimum adiabatic population, and peak Rydberg admixture are evaluated at the pulse duration selected at each parameter point under the constraint $\alpha_0\tau_{\rm gate}\leq160\pi$. Figure~\ref{fig:gate_map} shows how the gate performance changes across the parameter range as the system approaches the avoided crossing.

In \frefp{fig:gate_map}{a}, we see that the minimum population in the instantaneous dressed state $\ket{\widetilde{11} (t)}$ remains close to unity over most of the F\"orster-enhanced operating range and decreases only near the avoided crossing. The evolution is therefore adiabatic over the parameter range relevant for the gate. Since $\Omega(t)=0$ at the beginning and end of the pulse, the dressed state approaches $\ket{11}$ at both times, and adiabatic following returns the population to $\ket{11}$ at the end of the evolution. Accordingly, we find that the final return probabilities $|c_1|^2$ and $|c_{11}|^2$ remain close to one throughout this range.

In \frefp{fig:gate_map}{b}, the conditional phase error, $(\phi_{\rm CZ}-\pi)/\pi$ is shown. It vanishes within numerical accuracy at points where the target phase is reached. At small $\alpha_0$ and away from the avoided crossing, the evolution remains nearly adiabatic, but the dressed interaction is insufficient to accumulate a $\pi$ phase within the allowed duration. In this regime, F\"orster enhancement near the crossing is needed to reach the target phase. Increasing $\alpha_0$ allows the phase condition to be satisfied farther from the crossing.

The Bell state fidelity is shown in \frefp{fig:gate_map}{c}. The fidelity is high throughout the region where the target phase is reached and the population returns to the computational states. When the phase condition is satisfied exactly, the remaining deviation of $F_{\rm Bell}$ from unity is due to incomplete population return. A nonzero conditional phase error gives an additional contribution.

Finally, in \frefp{fig:gate_map}{d}, we show the peak Rydberg state population, $P_{\rm Ryd}^{\max}$. The state follows $\ket{\widetilde{11}(t)}$ where the evolution remains adiabatic, so the Rydberg population reflects the Rydberg admixture of this dressed state. Increasing $\alpha_0$ increases this admixture through the laser coupling, while approaching the avoided crossing increases the Rydberg pair-state contribution. Thus, a Bell state fidelity close to unity does not imply a small transient Rydberg population.

We identify the region in which the four gate conditions are simultaneously satisfied: $\phi_{\rm CZ}=\pi$, $P_{\rm adiab}^{\min}\geq0.999$, $F_{\rm Bell}\geq0.9999$, and $P_{\rm Ryd}^{\max}\leq0.11$. This region is shown by the red color in \frefp{fig:gate_map}{d}. Within this region, the operating point is chosen as the one with the smallest $\alpha_0\tau_{\rm gate}$. Since $\alpha_0\tau_{\rm gate}=\Omega_{\max}T$, this gives the shortest gate duration at fixed peak Rabi frequency among the points considered. The selected point has $P_{\rm Ryd}^{\max}\simeq0.11$ and is marked by the star in \frefp{fig:gate_map}{d}.

The bound $\alpha_0\tau_{\rm gate}\leq160\pi$ follows from the imposed limits on the peak Rabi frequency and pulse duration. In \aref{app:budget}, \frefp{fig:budget}{a}, shows that the minimum $P_{\rm Ryd}^{\max}$ among pulses satisfying the phase and adiabaticity conditions decreases over the range examined as $\alpha_0\tau_{\rm gate}$ increases. At fixed $\Omega_{\max}$, the required phase can therefore be accumulated with less Rydberg excitation by allowing a longer gate duration. However, increasing $\alpha_0\tau_{\rm gate}$ beyond $160\pi$ requires either $\Omega_{\max}/2\pi>8~{\rm MHz}$ or $T>10~\mu{\rm s}$. Within $\alpha_0\tau_{\rm gate}\leq160\pi$, we find no point in the present scan that satisfies the phase, adiabaticity, and fidelity conditions with $P_{\rm Ryd}^{\max}\leq0.10$. We therefore choose $P_{\rm Ryd}^{\max}\leq0.11$, for which these conditions can be satisfied.

For $\gamma=0$, the selected operating point is $\alpha_0\simeq0.2869$ and $\beta/\beta_{\rm res}\simeq0.9506$, with $\tau_{\rm gate}\simeq1613.49$ and $\alpha_0\tau_{\rm gate}\simeq463$. At this point, $P_{\rm Ryd}^{\max}\simeq0.11$ and $P_{\rm adiab}^{\min}\simeq0.99994$. During the pulse, the Rydberg population increases as the laser coupling is turned on and decreases to a negligible value as it is turned off. As described in \aref{app:gate_dynamics}, the conditional phase is accumulated predominantly near the maximum of the pulse, where the magnitude of the dressed interaction is largest, while the dynamics remains nearly adiabatic throughout. Next, we examine how the selected operating point changes as the F\"orster defect $\gamma$ is varied.

\begin{figure*}[htb]
\centering
\includegraphics[width=\textwidth]{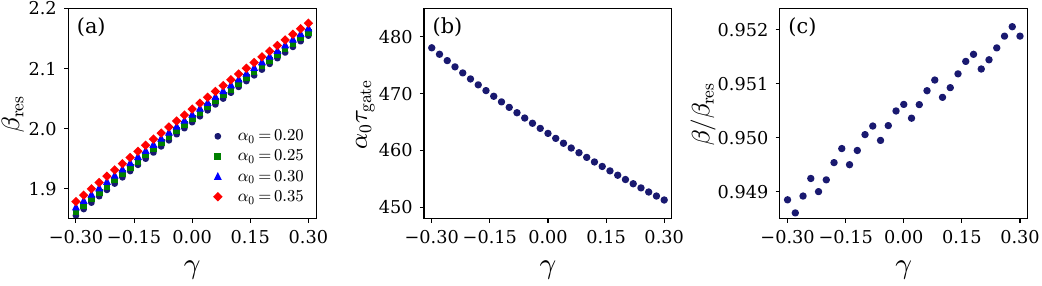}
\caption{
Dependence of the selected CZ-gate parameters on the relative F\"orster defect $\gamma=\delta_p/\Delta$. (a) Pair coupling $\beta_{\rm res}$ at the avoided crossing for $\alpha_0=0.20$, $0.25$, $0.30$, and $0.35$. (b) Minimum $\alpha_0\tau_{\rm gate}=\Omega_{\max}T$ among the points satisfying the gate conditions at each $\gamma$. (c) Corresponding normalized pair coupling $\beta/\beta_{\rm res}$.
}
\label{fig:gamma_optimized}
\end{figure*}
%

%%%%%%%%%%%%%%%%%%%%%%%%%%%%%%%%%%%%%%%%%%%%%%%%%%%%%%%%%%%%%%%%%%
\subsection{Dependence on the F\"orster defect}
\label{sec:gamma_dependence}
%%%%%%%%%%%%%%%%%%%%%%%%%%%%%%%%%%%%%%%%%%%%%%%%%%%%%%%%%%%%%%%%%%%
The relative F\"orster defect $\gamma$ sets the energy separation between $\ket{rr}$ and $\ket{pp'}$. Changing $\gamma$ shifts the position of the avoided crossing. We vary $\gamma$ from $-0.3$ to $+0.3$, for which the pair states remain nearly degenerate on the scale set by $\Delta$. For each $\alpha_0$ and $\gamma$, we determine $\beta_{\rm res}$ from the minimum of $E_{\rm gap}$ in \eref{eq:gap}, evaluated at the peak laser coupling. The pair coupling is then expressed as $\beta/\beta_{\rm res}$, so that $\beta/\beta_{\rm res}=1$ corresponds to the avoided crossing at those values of $\alpha_0$ and $\gamma$. As shown in \frefp{fig:gamma_optimized}{a}, $\beta_{\rm res}$ increases approximately linearly with $\gamma$, so the crossing shifts toward increasing pair coupling as $\gamma$ increases. Its dependence on $\alpha_0$ is weak over the range examined.

For each $\gamma$, we repeat the gate calculation of \sref{sec:gate_map}, apply the same four conditions, and select the point with the smallest $\alpha_0\tau_{\rm gate}$. Gates satisfying these conditions are found at every defect examined, with $P_{\rm Ryd}^{\max}\simeq0.11$ and $P_{\rm adiab}^{\min}>0.9999$. The required $\alpha_0\tau_{\rm gate}$ decreases by about $6\%$ between $\gamma=-0.3$ and $+0.3$, as shown in \frefp{fig:gamma_optimized}{b}. At fixed peak Rabi frequency, this gives the same fractional reduction in gate duration without an increase in the peak Rydberg population.

Although $\beta_{\rm res}$ shifts appreciably with the defect, the normalized pair coupling at the selected point changes only slightly, as shown in \frefp{fig:gamma_optimized}{c}. The dressing ratio also remains close to $\alpha_0=0.287$. The pair coupling selected for the gate therefore follows the shift of the avoided crossing, while the dressing strength and the relative position of the operating point remain nearly unchanged. The phase and adiabatic-return conditions can thus be maintained away from exact pair-state degeneracy by adjusting the pair coupling and pulse duration.

%%%%% Experimental %%%%%%%
Next, to relate the dimensionless operating points to the parameters used in experiments, we consider the $^{87}{\rm Rb}$ F\"orster channel
\[
58d_{3/2}+58d_{3/2} \rightarrow 60p_{1/2}+56f_{5/2}. 
\]
For this channel, the F\"orster defect is $\delta_p/2\pi=-6.92~{\rm MHz}$, and the dipole-dipole coefficient for the pair channel considered here is $C_3/h=3.200~{\rm GHz}\,\mu{\rm m}^3$ \cite{walker2008consequences,gaetan2009observation,vsibalic2017arc}. Since we consider positive laser detuning, this F\"orster defect corresponds to $\gamma<0$. Fixing $\delta_p$ determines the detuning for each negative $\gamma$ operating point through $\Delta/2\pi=(\delta_p/2\pi)/\gamma$.

\begin{figure}[htb]
\centering
\includegraphics[width=\columnwidth]{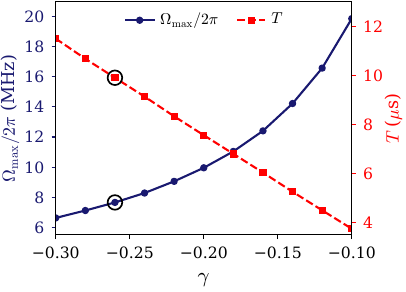}
\caption{
Parameters corresponding to the selected negative $\gamma$ operating points for the $^{87}{\rm Rb}$ F\"orster channel with $\delta_p/2\pi=-6.92~{\rm MHz}$. The peak Rabi frequency $\Omega_{\max}/2\pi$ (blue circles, left axis) and pulse duration $T$ (red squares, right axis) are plotted as functions of $\gamma$. For each operating point, the physical detuning is fixed by $\Delta=\delta_p/\gamma$. The open black circles mark $\gamma=-0.26$, the only value shown at which $T\leq10~\mu{\rm s}$ and $\Omega_{\max}/2\pi\leq8~{\rm MHz}$ both hold.
}
\label{fig:experimental}
\end{figure}

The peak Rabi frequency and pulse duration are obtained from $\Omega_{\max}=\alpha_0\Delta$ and $T=\tau_{\rm gate}/\Delta$, respectively, and their dependence on $\gamma$ is shown in \fref{fig:experimental}. The corresponding values of $\Delta$, $\Omega_{\max}$, $T$, and the interatomic distance $R$ for $\gamma=-0.30$ to $-0.20$ are given in Table~\ref{tab:rb_operating_points}.
\begin{table}[htb]
\caption{Detuning $\Delta$, peak Rabi frequency $\Omega_{\max}$, pulse duration $T$, and interatomic separation $R$ obtained from the selected operating point at each $\gamma$ for the $^{87}{\rm Rb}$ channel with $\delta_p/2\pi=-6.92~{\rm MHz}$ and $C_3/h=3.200~{\rm GHz}\,\mu{\rm m}^3$. The bold row satisfies both $T\leq10~\mu{\rm s}$ and $\Omega_{\max}/2\pi\leq8~{\rm MHz}$.}
\label{tab:rb_operating_points}
\begin{ruledtabular}
\begin{tabular}{*{5}{>{\color{black}}c}}
$\gamma$ & $\Delta/2\pi$ (MHz) & $\Omega_{\max}/2\pi$ (MHz) & $T$ ($\mu$s) & $R$ ($\mu$m) \\
\hline
$-0.30$ & 23.07 & 6.62 & 11.49 & 4.28 \\
$-0.28$ & 24.71 & 7.12 & 10.67 & 4.17 \\
$\mathbf{-0.26}$ & \textbf{26.62} & \textbf{7.65} & \textbf{9.90} & \textbf{4.06} \\
$-0.24$ & 28.83 & 8.28 & 9.13 & 3.95 \\
$-0.22$ & 31.45 & 9.05 & 8.33 & 3.83 \\
$-0.20$ & 34.60 & 9.95 & 7.56 & 3.70 \\
\end{tabular}
\end{ruledtabular}
\end{table}
Since $\delta_p$ is fixed by the chosen F\"orster channel, varying $\gamma$ changes the laser detuning. A smaller $|\gamma|$ requires a larger $\Delta$. For the selected operating point at each $\gamma$, the peak Rabi frequency therefore increases and the pulse duration decreases as shown in \fref{fig:experimental}.  For the nonzero F\"orster defect, $\gamma\to0$ requires $\Delta\to\infty$ and does not make the bare pair states degenerate. The $\gamma=0$ case considered in \sref{sec:gate_map} instead corresponds to $\delta_p=0$ at finite detuning.

The condition $\alpha_0\tau_{\rm gate} \le 160 \pi$ constrains the product $\Omega_{\max}T$, not the two physical quantities separately. We therefore check $\Omega_{\max}/2\pi\leq8~{\rm MHz}$ and $T\leq10~\mu{\rm s}$ after setting $\Delta=\delta_p/\gamma$ for the chosen channel. Among the values in Table~\ref{tab:rb_operating_points}, both conditions are satisfied at $\gamma=-0.26$. At this point, $\alpha_0\simeq0.2875$ and $\beta/\beta_{\rm res}\simeq0.9489$, giving $\Delta/2\pi\simeq26.62~{\rm MHz}$, $\Omega_{\max}/2\pi\simeq7.65~{\rm MHz}$, $T\simeq9.90~\mu{\rm s}$, and $B/2\pi\simeq47.7~{\rm MHz}$.

For the pair channel considered here, $B/2\pi=(C_3/h)/R^3$, giving an interatomic separation $R\simeq4.06~\mu{\rm m}$. The selected gate has $P_{\rm adiab}^{\min}\simeq0.99994$ and $P_{\rm Ryd}^{\max}\simeq0.11$. 

%%%%%%%%%%%%%%%%%%%%%%%%%%%%%%%%%%%%%%%%%%%%%%%%%%%
\section{Finite-lifetime error}
\label{sec:decay}
%%%%%%%%%%%%%%%%%%%%%%%%%%%%%%%%%%%%%%%%%%%%%%%%%%%
The coherent analysis above assumes that the Rydberg states do not decay during the pulse. Since the Rydberg states have finite lifetimes, their spontaneous decay introduces an additional contribution to the gate error. We include this effect using effective non-Hermitian Hamiltonians \cite{keating2015robust,mitra2023neutral},
\begin{equation}
H_{11}^{\rm eff}(t)
=
H_{11}(t)
-\frac{i\hbar\Gamma}{2}\ket{R}\bra{R}
-i\hbar\Gamma\ket{rr}\bra{rr}
-i\hbar\Gamma\ket{pp'}\bra{pp'} ,
\label{eq:Hdecay_two}
\end{equation}
and
\begin{equation}
H_1^{\rm eff}(t)
=
H_1(t)
-\frac{i\hbar\Gamma}{2}\ket{r}\bra{r}.
\label{eq:Hdecay_single}
\end{equation}
Here, $\Gamma$ is the decay rate of a single Rydberg excitation which is taken to be the same for the $r$, $p$, and $p'$ states.
The state $\ket{R}$ contains one Rydberg excitation, whereas
$\ket{rr}$ and $\ket{pp'}$ each contain two. This gives the corresponding non-Hermitian decay terms in Eq.~\eqref{eq:Hdecay_two}.

\begin{figure}[t]
    \centering
    \includegraphics[width=\columnwidth]{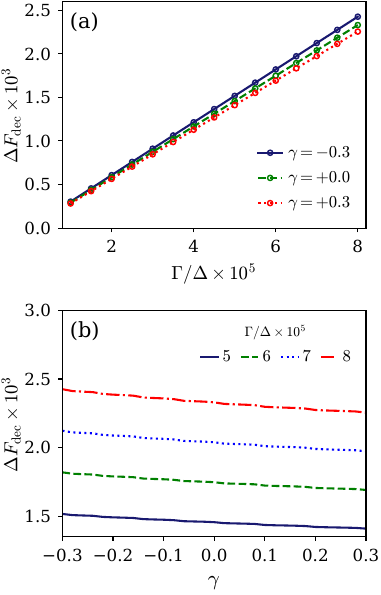}
    \caption{
    Finite-lifetime contribution to the Bell-state fidelity loss for the selected CZ gates. (a) $\Delta F_{\rm dec}$ as a function of $\Gamma/\Delta$ for $\gamma=-0.3$, $0$, and $+0.3$. (b) $\Delta F_{\rm dec}$ as a function of $\gamma$ for $\Gamma/\Delta=(5,6,7,8)\times10^{-5}$. For each $\gamma$, the decay calculation uses the corresponding lossless operating point without reoptimizing the gate parameters.
    }
    \label{fig:decay}
\end{figure}

These terms describe decay through the loss of norm during evolution. We do not renormalize the final state when evaluating the fidelity, so this loss contributes directly to the gate error. This treatment does not include repopulation of lower atomic states or branching among different decay channels \cite{petrosyan2017high}. A full description of these processes would require a master-equation treatment. We therefore use the non-Hermitian model to estimate the error associated with the finite Rydberg lifetimes.

For each value of the relative F\"orster defect $\gamma$, we use the operating point selected from the corresponding lossless calculation and then include Rydberg decay without reoptimizing the gate parameters. We denote the resulting Bell-state fidelity by $F_{\rm Bell}^{(\rm dec)}$ and define
\begin{equation}
\Delta F_{\rm dec}
=
F_{\rm Bell}^{(0)}
-
F_{\rm Bell}^{(\rm dec)},
\label{eq:DeltaFdec}
\end{equation}
where $F_{\rm Bell}^{(0)}$ is the lossless fidelity at the same operating point. Thus, $\Delta F_{\rm dec}$ gives the change in fidelity due to Rydberg state decay at fixed gate parameters.

Since the gate dynamics is expressed in units of $\Delta$, the decay rate enters through the dimensionless ratio $\Gamma/\Delta$. We vary $\Gamma/\Delta$ from $10^{-5}$ to $8\times10^{-5}$, which covers the values corresponding to Rydberg lifetimes in the range $100$-$300~\mu\mathrm{s}$ for $\Delta/2\pi=20$-$40~\mathrm{MHz}$, relevant to Rydberg-dressing gates \cite{mitra2023neutral}.

In \frefp{fig:decay}{a}, we show the finite-lifetime error as $\Gamma/\Delta$ is varied for the selected gates at $\gamma=-0.3$, $0$, and $+0.3$. For all three relative F\"orster defects, $\Delta F_{\rm dec}$ increases nearly linearly with $\Gamma/\Delta$. Over this range, the loss remains small, and increasing $\Gamma/\Delta$ increases the population lost through Rydberg decay during the pulse. The similar slopes of the three curves indicate that the selected gates have comparable sensitivity to Rydberg decay over this range of $\gamma$.

The dependence on the F\"orster defect is shown in \frefp{fig:decay}{b}. At fixed $\Gamma/\Delta$, $\Delta F_{\rm dec}$ decreases slightly as $\gamma$ is varied from $-0.3$ to $+0.3$. For example, at $\Gamma/\Delta=5\times10^{-5}$, $\Delta F_{\rm dec}$ decreases from approximately $1.52\times10^{-3}$ to $1.41\times10^{-3}$. Although $P_{\rm Ryd}^{\max}\simeq0.11$ at every selected point, the Rydberg population evolves differently during the pulse for different $\gamma$. Since the decay accumulates over the entire pulse, the same peak population can therefore give different finite-lifetime errors. This distinction is also seen in \aref{app:budget}, \fref{fig:budget}. There, increasing $\alpha_0\tau_{\rm gate}$ reduces the minimum peak Rydberg population, but the accumulated decay error increases at fixed $\Gamma/\Delta$ over the range examined. The peak Rydberg population alone therefore does not determine the finite-lifetime error.

\begin{figure*}[htb]
\centering
\includegraphics[width=\textwidth]{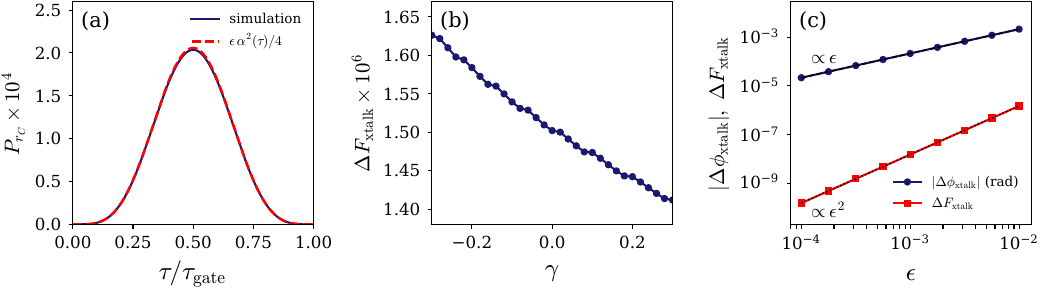}
\caption{
Spectator-induced effects on the addressed CZ gate. Unless varied explicitly, the spectator parameters are $\epsilon=10^{-2}$, $v_0=0.3$, $\zeta=1$, and $\eta=1$. (a) Spectator Rydberg state population during the pulse for the addressed input $\ket{01}$. The numerical result is compared with the weak-dressing expression $\epsilon\alpha^2(\tau)/4$. (b) Spectator-induced Bell-state fidelity loss $\Delta F_{\rm xtalk}$ as a function of the F\"orster defect $\gamma$, using the corresponding selected gate operating point for each $\gamma$. (c) Conditional phase change $|\Delta\phi_{\rm xtalk}|$ and fidelity loss $\Delta F_{\rm xtalk}$ as functions of the leaked intensity $\epsilon$ at $\gamma=0$. The reference lines indicate the dependences $|\Delta\phi_{\rm xtalk}|\propto\epsilon$ and $\Delta F_{\rm xtalk}\propto\epsilon^2$.
}
\label{fig:crosstalk}
\end{figure*}
%

%%%%%%%%%%%%%%%%%%%%%%%%%%%%%%%%%%%%%%%%%%%%%%%%%%%
\section{Spectator-induced crosstalk}
%%%%%%%%%%%%%%%%%%%%%%%%%%%%%%%%%%%%%%%%%%%%%%%%%%%
\label{sec:crosstalk}

The finite-lifetime error considered above arises from the transient Rydberg population of the two addressed atoms, $A$ and $B$. A separate source of error is imperfect spatial addressing, for which a fraction of the laser intensity driving both $A$ and $B$ also reaches a neighboring spectator atom $C$. We therefore include a third atom $C$, initially in the state $\ket{1}_C$, and calculate the resulting changes in the accumulated phase during the pulse and in the fidelity of the addressed CZ gate. This contribution is referred to as spectator-induced crosstalk.

The spectator atom is described in the basis $\{ \ket{1}_C,\ket{r}_C\}$. We denote by $\epsilon$ the fraction of the laser intensity reaching atom $C$ relative to the intensity at the gate atoms. Its Rabi frequency is
\begin{equation}
\Omega_C(t)
=
\sqrt{\epsilon}\,\Omega(t),
\qquad
\alpha_C(\tau)
=
\sqrt{\epsilon}\,\alpha(\tau).
\label{eq:OmegaC}
\end{equation}
The interaction of a Rydberg excited spectator with gate atoms $A$ and $B$ is represented by the van der Waals shifts $V_{AC}$ and $V_{BC}$, respectively. The resulting three-atom dynamics are governed by \cite{warttmann2026suppressing}
\begin{equation}
H_{ABC}(t)
=
H_{AB}(t)\otimes I_C
+
I_{AB}\otimes H_C(t)
+
H_{\rm int}^{(C)} .
\label{eq:HABC}
\end{equation}
Here, $H_{AB}(t)$ describes the driven gate atoms, including the F\"orster coupling $\ket{rr}\leftrightarrow\ket{pp'}$, and $H_C(t)$ the spectator driven with $\Omega_C(t)$.
The term $H_{\rm int}^{(C)}$ accounts for the interaction-induced energy shifts when the spectator occupies $\ket{r}_C$. Since $V_{AC}$ and $V_{BC}$ need not be equal, states containing a Rydberg excitation on atom $A$ or atom $B$ are retained separately. The resulting three-atom basis and the corresponding Hamiltonian matrix elements are given in \aref{app:crosstalk} 1. Here, the spectator interaction is described by an effective shift model, in which additional Rydberg pair channels involving the spectator are not included.

In the dimensionless form introduced in \sref{sec:dimensionless}, we define
\begin{equation}
v_0=\frac{V_{BC}}{\Delta},
\qquad
\zeta=\frac{V_{AC}}{V_{BC}},
\label{eq:v0_zeta}
\end{equation}
where $v_0$ is the spectator interaction with atom $B$ relative to the detuning, while $\zeta$ gives the interaction with atom $A$ relative to that with atom $B$. When both gate atoms are in the Rydberg state, the two spectator interactions add, giving a shift $(1+\zeta)V_{BC}$ of $\ket{rr}$. The spectator-induced shift of $\ket{pp'}$ can differ from that of $\ket{rr}$. We account for this difference by introducing the factor $\eta$, so that the shift of $\ket{pp'}$ is $\eta(1+\zeta)V_{BC}$.

We first calculate the spectator Rydberg population
$P_{r_C}(\tau)$ to determine whether the weak spectator drive leaves residual Rydberg excitation at the end of the pulse. Even when the final population is negligible, the transient excitation of atom $C$ shifts the Rydberg states of the addressed pair and changes their accumulated phases. Since the Rydberg admixture depends on the computational input $\ket{ij}$, these phase shifts are generally state dependent. Moreover, when $V_{AC}\neq V_{BC}$, the phases acquired from $\ket{01}$ and $\ket{10}$ need not be equal. The conditional phase is determined from
\begin{equation}
\phi^{\rm (xtalk)}_{\rm cond}
=
\phi_{11}-\phi_{10}-\phi_{01}+\phi_{00},
\label{eq:phi_cond_xtalk}
\end{equation}
where $\phi_{ij}$ is the phase of the return amplitude for the initial state $\ket{ij}\ket{1}_C$. The spectator-induced change in the conditional phase is then
\begin{equation}
\Delta\phi_{\rm xtalk}
=
\phi^{\rm (xtalk)}_{\rm cond}
-
\phi_{\rm cond}^{(0)}.
\label{eq:Deltaphi_xtalk}
\end{equation}
The superscript $(0)$ denotes the spectator-free two-atom gate,
equivalently obtained from the three-atom model by setting
$V_{AC}=V_{BC}=0$.

To quantify the effect of the spectator on the addressed gate, we calculate the Bell-state fidelity for the initial state $\ket{++}_{AB}\ket{1}_C$. After the evolution, atom $C$ is traced out, and the same local $Z$ correction used for the spectator free gate in Sec.~\ref{sec:pulse_metrics} is applied to atoms $A$ and $B$. The resulting fidelity includes both spectator-induced phase shifts and any population that does not return to the $AB$ computational states. The details of this calculation are given in \aref{app:crosstalk} 2. The corresponding fidelity loss relative to the spectator-free gate is
\begin{equation}
\Delta F_{\rm xtalk}
=
F_{\rm Bell}^{(0)}
-
F_{\rm Bell}^{(\rm xtalk)}.
\label{eq:DeltaF_xtalk}
\end{equation}

We vary the leakage fraction $\epsilon$ from $10^{-4}$ to $10^{-2}$, corresponding to $0.01\%$-$1\%$ intensity leakage. The spectator interaction $v_0$ is varied from $0.01$ to $0.5$, remaining below the addressed-pair coupling for the gate parameters considered here. The interaction ratio $\zeta$ is varied from $0.1$ to $1$, covering asymmetric spectator interactions with atoms $A$ and $B$ up to the case of equal interactions. The factor $\eta$ is varied from $0.1$ to $2$, allowing the spectator-induced shift of $\ket{pp'}$ to range from smaller than that of $\ket{rr}$ to twice its value.

The dependence of $\Delta F_{\rm xtalk}$ on $\epsilon$, $v_0$, $\zeta$, and $\eta$ over the ranges considered above is shown in \aref{app:crosstalk} 3. For the description here, we take $\epsilon=10^{-2}$, $v_0=0.3$, $\zeta=1$, and $\eta=1$. We first consider the input $\ket{01}$, for which only atom $B$ of the addressed pair is driven. This provides the single-atom case for examining the excitation of the spectator by the leaked laser field. In \frefp{fig:crosstalk}{a} the spectator Rydberg population during the pulse is shown. In the weak-dressing limit, the spectator Rydberg population is
\begin{equation}
P_{r_C}(\tau)
\simeq
\frac{\epsilon\alpha^2(\tau)}{4},
\label{eq:PrC_weak}
\end{equation}
since the spectator Rabi frequency is
$\Omega_C=\sqrt{\epsilon}\,\Omega$. The numerical result
follows this expression closely. For $\epsilon=10^{-2}$, $P_{r_C}$ reaches approximately $2.0\times10^{-4}$ near the
pulse maximum and returns to a negligible value at the end
of the pulse. The leaked field therefore produces only a
small transient Rydberg excitation of the spectator.

This behavior differs from the single pulse protocol of \rref{warttmann2026suppressing}, in which both the Rabi frequency and detuning are varied during the gate and the pulse leaves a finite population in the spectator Rydberg state. In the present fixed detuning protocol, the spectator returns almost completely to $\ket{1}_C$, with negligible residual Rydberg population at the end of the pulse.

In \frefp{fig:crosstalk}{b}, we show how the crosstalk-induced fidelity loss changes with the F\"orster defect. For each value of $\gamma$, we use the corresponding gate operating point determined in \sref{sec:gamma_dependence}. The value of $\Delta F_{\rm xtalk}$ over the range $\gamma=-0.3$ to $0.3$ remains
of order $10^{-6}$, changing from approximately $1.63\times10^{-6}$ to $1.41\times10^{-6}$. Thus, changing
the F\"orster defect and the corresponding gate operating
point has only a small effect on the spectator-induced fidelity loss. For $\epsilon=10^{-2}$, this contribution is about three orders of magnitude smaller than the finite-lifetime error obtained at $\Gamma/\Delta=5.0\times10^{-5}$.

The dependence on the leaked intensity is shown in \frefp{fig:crosstalk}{c}. In the weak leakage regime, $\Omega_C\propto\sqrt{\epsilon}$ so the spectator Rydberg population scales as $P_{r_C}\propto\epsilon$. The spectator-gate interaction acts only while atom $C$ has a finite Rydberg population, so the resulting conditional phase change is proportional to $\epsilon$ for fixed $v_0$. Since the population return remains nearly complete, the fidelity loss is dominated by the spectator induced phase changes and scales quadratically with $\epsilon$. The numerical results follow
\begin{equation}
|\Delta\phi_{\rm xtalk}|
\propto
\epsilon,
\qquad
\Delta F_{\rm xtalk}
\propto
\epsilon^2.
\label{eq:xtalk_scaling}
\end{equation}
For the parameters used in Fig.~\ref{fig:crosstalk}, the conditional phase change is approximately $|\Delta\phi_{\rm xtalk}|=2.2~{\rm mrad}$. The population remaining outside the $AB$ qubit states at the end of the pulse is of order $10^{-11}$, while the Bell-state fidelity is reduced by $\Delta F_{\rm xtalk}\simeq1.5\times10^{-6}$. The fidelity loss, therefore, comes almost entirely from the additional phase accumulated during the pulse rather than from the incomplete population return.

\section{Conclusion}
In conclusion, we have studied the effect of a near-degenerate F\"orster pair channel in the adiabatic Rydberg-dressing CZ gate. By including the dipole-dipole coupling to the near-degenerate pair channel in the time-dependent gate dynamics, we identify a F\"orster-enhanced operating range in which the required conditional phase is accumulated with adiabatic return and small transient Rydberg population. In this range, the pair-state coupling enhances the dressed interaction before the sharp reduction of the dressed state gap and growth of the Rydberg admixture near the avoided crossing.

We show that the gate operating point within this range is determined by the balance between conditional phase accumulation, adiabatic following, and transient Rydberg population. Among the parameters that produce the required conditional phase at fixed peak Rabi frequency and pulse duration, a small dressing ratio requires operation close to the avoided crossing, where the Rydberg admixture increases. At large dressing ratios, the dressing field itself increases the Rydberg population. The competition between these two effects results in a minimum in the peak Rydberg state population. We select the operating point with the smallest $\alpha_0\tau_{\rm gate}$ that satisfies the phase, adiabaticity, and Bell-state fidelity conditions while limiting the peak Rydberg population to $P_{\rm Ryd}^{\max}\leq0.11$. At fixed peak Rabi frequency, this gives the shortest gate within the pulse and parameter ranges examined.

We further analyze how the F\"orster defect affects the selected operating point within the F\"orster-enhanced range. Changing the defect shifts the position of the avoided crossing, but the position of the operating point relative to the crossing, characterized by $\beta/\beta_{\rm res}$, changes only slightly. The selected dressing ratio $\alpha_0$ remains nearly the same, and the gate duration at fixed peak Rabi frequency depends only weakly on the defect over the range considered. To relate these results to a specific atomic system, we then consider a rubidium F\"orster channel and determine the corresponding Rabi frequency, pulse duration, and interatomic separation for a selected gate that satisfies the imposed bounds on peak Rabi frequency and pulse duration.

Finally, we examine the effects of the finite Rydberg lifetime and spectator-induced crosstalk on the selected gate. For a given decay, we find that increasing $\alpha_0\tau_{\rm gate}$ reduces the minimum peak Rydberg-state population but increases the finite-lifetime error of the corresponding gates. The decay error depends on the Rydberg population throughout the pulse, not only on its maximum. These results show that the peak population alone is insufficient for selecting the gate with minimum decay error. For spectator-induced crosstalk, the weakly driven spectator has negligible Rydberg population at the end of the pulse, but its transient excitation shifts the Rydberg-state energies of the gate atoms and changes both the single-qubit and conditional phases. The resulting crosstalk error remains below the finite-lifetime contribution within the effective shift model and for the parameter range considered here.

An important next step is to include atomic motion. The present calculations assume a fixed interatomic separation, whereas the position dependence of the dipole-dipole coupling makes the pair coupling sensitive to motion. Near the avoided crossing, variations in the interatomic separation can modify both the conditional interaction and the dressed state energy gap and thereby couple the internal gate dynamics to the atomic motion. A dynamical treatment should therefore include the atomic wave packets, Doppler shifts, spatial variation of the dressing field, and interaction-induced forces during the pulse. The spectator model can also be extended beyond the diagonal interaction shifts considered here by including additional Rydberg pair channels involving the spectator atoms. Such channels may introduce additional couplings during the transient spectator excitation and modify the resulting crosstalk. Extending the analysis to larger arrays, with several neighboring atoms and simultaneous gate operations, will then be important for determining how the F\"orster-enhanced gate performs in a multiqubit system and for developing pulse protocols that are robust against atomic motion, laser fluctuations, and spectator interactions.

%%%%%%%%%%%%%%%%%%%%%%%%%%%%%%%%%%%%%%%%%%%%%%%%%%%
\section{Acknowledgment}
%%%%%%%%%%%%%%%%%%%%%%%%%%%%%%%%%%%%%%%%%%%%%%%%%%%

This work is supported by the Department of Science and Technology, Government of India, under the National Quantum Mission (NQM). D.T. acknowledges financial support from the National Quantum Mission (NQM) and thanks Ishitwa Kumar Das for helpful discussions.

%%%%%%%%%%%%%%%%%%%%%%%%%%%%%%%%%%%%%%%%%%%%%%%%%%%
\appendix
%%%%%%%%%%%%%%%%%%%%%%%%%%%%%%%%%%%%%%%%%%%%%%%%%%%

%%%%%%%%%%%%%%%%%%%%%%%%%%%%%%%%%%%%%%%%%%%%%%%%%%%
\section{Time evolution and additional gate analysis}
%%%%%%%%%%%%%%%%%%%%%%%%%%%%%%%%%%%%%%%%%%%%%%%%%%%
\label{app:eom}

The time-dependent calculations in \sref{sec:cz_dynamics} are performed in the bare-state basis of the single and two-atom Hamiltonian introduced in \sref{sec:dressed_mechanism}. Here, we provide the equations used for the numerical simulation and the procedure used to follow the instantaneous dressed state connected to $\ket{11}$. We then use these quantities to examine the complete evolution at the selected operating point and the dependence of the gate on the pulse duration. 

\begin{figure*}[htb]
\centering
\includegraphics[width=0.98\textwidth]{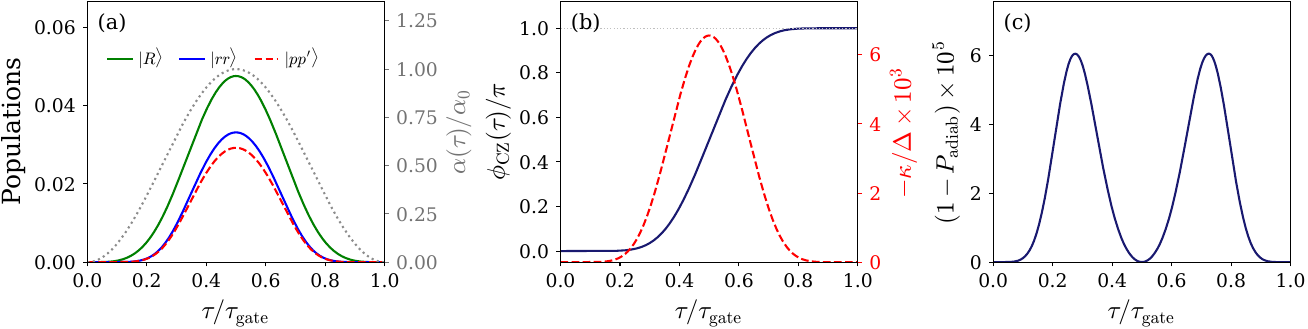}
\caption{ Time-dependent dynamics at the selected $\gamma=0$ operating point, $\alpha_0=0.2869$ and $\beta/\beta_{\rm res}=0.9506$. (a) Populations of $\ket{R}$, $\ket{rr}$, and $\ket{pp'}$ starting from $\ket{11}$. The gray dotted curve shows the pulse envelope $\alpha(\tau)/\alpha_0$. (b) Accumulated conditional phase $\phi_{\rm CZ}(\tau)/\pi$ together with the instantaneous dressed interaction $-\kappa(\tau)/\Delta$. (c) Deviation $1-P_{\rm adiab}(\tau)$ from the instantaneous $\ket{11}$-connected dressed state.
}
\label{fig:gate_dynamics}
\end{figure*}
%

%%%%%%%%%%%%%%%%%%%%%%%%%%%%%%%%%%%%%%%%%%%%%%%%%%%
\subsection{Time evolution in bare basis}
\label{app:eom_1}
%%%%%%%%%%%%%%%%%%%%%%%%%%%%%%%%%%%%%%%%%%%%%%%%%%%

We use the dimensionless time $\tau=\Delta t$ introduced in \sref{sec:dimensionless} and define the dimensionless Hamiltonians $\mathcal H_1=H_1/(\hbar\Delta)$ and $\mathcal H_{11}=H_{11}/(\hbar\Delta)$. For the single atom problem, we expand the total wave-function in the basis $\{\ket{1},\ket{r}\}$ as $\ket{\psi_1(\tau)}=\sum_n c_n^{(1)}(\tau)\ket{n}$, similarly for the two-atom problem, we write $\ket{\psi_{11}(\tau)}=\sum_n c_n(\tau)\ket{n}$ in the basis $\{\ket{11},\ket{R},\ket{rr},\ket{pp'}\}$. The Schr\"odinger equations of motion are then
\begin{align}
i\frac{d c_n^{(1)}}{d\tau}
&=
\sum_m [\mathcal H_1(\tau)]_{nm}c_m^{(1)},
\nonumber\\
i\frac{d c_n}{d\tau}
&=
\sum_m [\mathcal H_{11}(\tau)]_{nm}c_m,
\label{eq:tdse}
\end{align}
with the initial conditions $c_n^{(1)}(0)=\delta_{n,1}$ and $c_n(0)=\delta_{n,11}$. The detuning $\Delta$, pair coupling $B$, and F\"orster defect $\delta_p$ are held fixed during the pulse, while the dimensionless laser coupling is $\alpha(\tau)=\alpha_0\sin^2(\pi\tau/\tau_{\rm gate})$. The final dimensionless time is $\tau_{\rm gate}=\Delta T$, so the return amplitudes are
\begin{equation}
c_1=c_1^{(1)}(\tau_{\rm gate}),
\qquad
c_{11}=c_{11}(\tau_{\rm gate}).
\label{eq:return_amplitudes_app}
\end{equation}
The instantaneous dressed basis introduced next is used only to analyze the evolution and to evaluate the adiabatic population defined in \eref{eq:Padiab}.

%%%%%%%%%%%%%%%%%%%%%%%%%%%%%%%%%%%%%%%%%%%%%%%%%%%
\subsection{Instantaneous dressed state tracking}
%%%%%%%%%%%%%%%%%%%%%%%%%%%%%%%%%%%%%%%%%%%%%%%%%%%
At each dimensionless time $\tau$, the instantaneous eigenstates can equivalently be obtained from the two-atom Hamiltonian,
\begin{equation}
\mathcal H_{11}(\tau)\ket{\varphi_k(\tau)}
=
\varepsilon_k(\tau)\ket{\varphi_k(\tau)},
\label{eq:adiabatic_basis_app}
\end{equation}
where $\varepsilon_k(\tau)=E_k(\tau)/(\hbar\Delta)$ are the dimensionless eigenenergies. We write $\ket{\varphi_k(\tau)}=\sum_n f_{kn}(\tau)\ket{n}$ with $f_{kn}(\tau)=\langle n|\varphi_k(\tau)\rangle$.

The instantaneous eigenstate relevant to the gate is the one continuously connected to $\ket{11}$. At the beginning of the pulse, where $\alpha(0)=0$, we identify this state from its overlap with $\ket{11}$. At each subsequent value of $\tau$, we select the eigenvector having the largest overlap with the eigenvector selected at the preceding step. This continuity criterion avoids a change of label when the bare-state composition of neighboring dressed states varies during the pulse. The resulting state is the $\ket{\widetilde{11}(\tau)}$ used in \sref{sec:pulse_metrics}.

The relation between the instantaneous state and nonadiabatic transitions can be seen by expanding the wave function as $\ket{\psi_{11}(\tau)}=\sum_k \widetilde c_k(\tau)\ket{\varphi_k(\tau)}$. In dimensionless time, the corresponding amplitudes obey
\begin{equation}
i\frac{d\widetilde c_n}{d\tau}
=
\varepsilon_n(\tau)\widetilde c_n
-i\sum_k d_{nk}(\tau)\widetilde c_k,
\label{eq:adiabatic_eom_app}
\end{equation}
where $d_{nk}(\tau)=\langle\varphi_n(\tau)|\partial_\tau\varphi_k(\tau)\rangle$ is the nonadiabatic coupling \cite{pant2024adiabaticity}. For $n\neq k$, differentiating Eq.~\eqref{eq:adiabatic_basis_app} gives
\begin{equation}
d_{nk}(\tau)
=
\frac{\langle\varphi_n(\tau)|\partial_\tau\mathcal H_{11}(\tau)|\varphi_k(\tau)\rangle}
{\varepsilon_k(\tau)-\varepsilon_n(\tau)}
\label{eq:nonadiabatic_coupling}
\end{equation}
which relates the static gap discussed in \sref{sec:dressed_mechanism} to the finite-time dynamics. A smaller separation in energy of the neighboring dressed states increases the nonadiabatic coupling, whereas a slower variation of the laser coupling suppresses it. Since $\Delta$, $B$, and $\delta_p$ remain fixed during the pulse, the time dependence of $H_{11}$ enters only through $\alpha(\tau)$ so the numerator of \eref{eq:nonadiabatic_coupling} is controlled by $d\alpha/d\tau$.

We do not propagate \eref{eq:adiabatic_eom_app} directly. Instead, the state obtained from the evolution in bare basis in \eref{eq:tdse} is projected onto the instantaneous eigenstates,
\begin{equation}
\widetilde c_k(\tau)
=
\langle\varphi_k(\tau)|\psi_{11}(\tau)\rangle
=
\sum_n f_{kn}^{*}(\tau)c_n(\tau).
\label{eq:adiabatic_projection_app}
\end{equation}
The adiabatic population of the tracked state is then
\begin{equation}
P_{\rm adiab}(\tau)
=
\left|\langle\widetilde{11}(\tau)|\psi_{11}(\tau)\rangle\right|^2,
\qquad
P_{\rm adiab}^{\min}=\min_{\tau} P_{\rm adiab}(\tau),
\label{eq:Padiab_app}
\end{equation}
defined in \eref{eq:Padiab}.

%%%%%%%%%%%%%%%%%%%%%%%%%%%%%%%%%%%%%%%%%%%%%%%%%%%
\subsection{Gate dynamics at operating point}
%%%%%%%%%%%%%%%%%%%%%%%%%%%%%%%%%%%%%%%%%%%%%%%%%%%
\label{app:gate_dynamics}

The parameter scan in \sref{sec:gate_map} identifies the operating point from the conditional phase, population return, adiabatic following, and peak Rydberg admixture. Figure~\ref{fig:gate_dynamics} shows how these quantities evolve during the pulse at the selected $\gamma=0$ operating point, $\alpha_0=0.2869$ and $\beta/\beta_{\rm res}=0.9506$.

As the pulse is turned on, population is admixed into $\ket{R}$, $\ket{rr}$, and $\ket{pp'}$, as shown in \frefp{fig:gate_dynamics}{a}. These Rydberg state populations are transient and decrease to negligible values as the pulse is turned off, while the population returns to $\ket{11}$ at the end of the evolution, consistent with the near-unity return probability found in \sref{sec:gate_map}. The Rydberg populations are largest near the center of the pulse, where the dressing is maximum.

The conditional phase in \frefp{fig:gate_dynamics}{b} is accumulated mainly over the same central part of the pulse. The magnitude of the dressed interaction is largest in this interval and therefore gives the dominant contribution to the phase integral in \eref{eq:conditional_phase}. By the end of the pulse, the conditional phase reaches $\pi$ while the transient Rydberg populations have returned to their initial values. The gate therefore acquires the required two-qubit phase through the dressed interaction without leaving residual Rydberg population at the end of the evolution.

Figure~\ref{fig:gate_dynamics}(c) shows the deviation from the tracked dressed state during the same pulse. The deviation remains small throughout the evolution and has two maxima on the rising and falling parts of the pulse. This behavior follows from \eref{eq:nonadiabatic_coupling}. Although the dressed state gap is smallest near the pulse maximum, the pulse envelope has zero slope at this point, and $d\alpha/d\tau=0$ at $\tau=\tau_{\rm gate}/2$. The nonadiabatic coupling is therefore suppressed at the center of the pulse. On either side of the maximum, $d\alpha/d\tau$ is nonzero resulting in small deviation from the instantaneous dressed state.

%%%%%%%%%%%%%%%%%%%%%%%%%%%%%%%%%%%%%%%%%%%%%%%%%%%
\subsection{Dependence on the pulse duration}
\label{app:budget}
%%%%%%%%%%%%%%%%%%%%%%%%%%%%%%%%%%%%%%%%%%%%%%%%%%%

In \sref{sec:gate_map}, we select the operating point at a given Rabi frequency as the smallest $\alpha_0\tau_{\rm gate}$ among gates satisfying $\phi_{\rm CZ}=\pi$, $P_{\rm adiab}^{\min}\geq0.999$, $F_{\rm Bell}\geq0.9999$, and $P_{\rm Ryd}^{\max}\leq0.11$. To examine how this choice affects the Rydberg population and the decay error, we repeat the same calculations over the range $302\leq\alpha_0\tau_{\rm gate}\leq754$, which correspond to $T = 6-15~\mu s$ at $\Omega_{\rm max}/2\pi = 8$ MHz. For each $\alpha_0\tau_{\rm gate}$ and $\gamma$, we determine the curve satisfying $\phi_{\rm CZ}=\pi$ and select the point with the smallest $P_{\rm Ryd}^{\max}$ among the points with $P_{\rm adiab}^{\min}\geq0.999$ and $0.20\leq\alpha_0\leq0.35$. We call a gate adiabatic when $P_{\rm adiab}^{\min}\geq0.999$, with $P_{\rm adiab}^{\min}$ defined in Eq.~\eqref{eq:Padiab}. We then include decay at these lossless operating points using the non-Hermitian Hamiltonian of \sref{sec:decay}, with $\Gamma/\Delta=5\times10^{-5}$.
\begin{figure}[htb]
\centering
\includegraphics[width=\columnwidth]{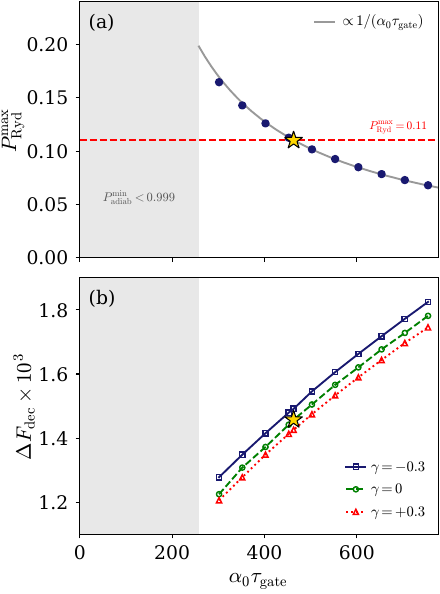}
\caption{
Dependence of the selected gate on $\alpha_0\tau_{\rm gate}=\Omega_{\max}T$. (a) Minimum peak Rydberg state population $P_{\rm Ryd}^{\max}$ at $\gamma=0$, obtained by selecting, at each $\alpha_0\tau_{\rm gate}$, the smallest $P_{\rm Ryd}^{\max}$ among points satisfying $\phi_{\rm CZ}=\pi$, $P_{\rm adiab}^{\min}\geq0.999$, and $0.20\leq\alpha_0\leq0.35$. The gray curve shows the $1/(\alpha_0\tau_{\rm gate})$ scaling. The horizontal dashed line marks $P_{\rm Ryd}^{\max}=0.11$. (b) Finite-lifetime contribution $\Delta F_{\rm dec}$ at the selected gates for $\gamma=-0.3$, $0$, and $+0.3$, with $\Gamma/\Delta=5\times10^{-5}$. The stars mark the $\gamma=0$ operating point used in the main analysis, at $\alpha_0\tau_{\rm gate}=463$. The shaded region denotes $\alpha_0\tau_{\rm gate}<258$, where $P_{\rm adiab}^{\min}<0.999$ for every point satisfying $\phi_{\rm CZ}=\pi$ in the examined range at $\gamma=0$.
}
\label{fig:budget}
\end{figure}
In \frefp{fig:budget}{a}, we show the minimum peak Rydberg admixture at each duration for $\gamma=0$. Shorter pulses require a larger $P_{\rm Ryd}^{\max}$ because the same conditional phase must be accumulated in less time. The required increase in the dressed interaction is obtained through stronger laser dressing, operation closer to the avoided crossing, or both. Stronger dressing increases the Rydberg population through the laser coupling, whereas proximity to the crossing increases the contribution of the Rydberg pair states. A longer pulse allows the required phase to be reached with less Rydberg admixture. Over the range considered, the selected values of $P_{\rm Ryd}^{\max}$ decrease approximately as $1/(\alpha_0\tau_{\rm gate})$.

The value $P_{\rm Ryd}^{\max}=0.11$ is reached at $\alpha_0\tau_{\rm gate}=463$. For shorter durations, the phase condition can still be satisfied over part of the parameter range, but the required Rydberg admixture exceeds this value. Below approximately $\alpha_0\tau_{\rm gate}<258$, $P_{\rm adiab}^{\min}<0.999$ for every point satisfying $\phi_{\rm CZ}=\pi$ in the examined range at $\gamma=0$, so we do not obtain an adiabatic CZ solution there. This lower limit therefore applies to the pulse and parameter range considered here, rather than to CZ gates in general.

In \frefp{fig:budget}{b}, we show the finite-lifetime error for $\gamma=-0.3$, $0$, and $+0.3$. In contrast to the peak Rydberg admixture, $\Delta F_{\rm dec}$ increases with the $\alpha_0\tau_{\rm gate}$. For $\gamma=0$, it rises from approximately $1.22\times10^{-3}$ at $\alpha_0\tau_{\rm gate}=302$ to $1.78\times10^{-3}$ at $\alpha_0\tau_{\rm gate}=754$. The positive defect gate has the smallest decay error throughout this range, while the negative-defect gate has the largest. This weak dependence on $\gamma$ is consistent with the results discussed in \sref{sec:decay}.

The different dependences of $P_{\rm Ryd}^{\max}$ and $\Delta F_{\rm dec}$ arise because decay depends on the Rydberg population throughout the pulse rather than only on its maximum. In \esref{eq:Hdecay_two}{eq:Hdecay_single}, each Rydberg excitation contributes to the norm loss, so the two-excitation states contribute with twice the rate of a singly excited state. At fixed $\Gamma/\Delta$, the accumulated loss depends on these populations over the full dimensionless duration $\tau_{\rm gate}=\Delta T$. The populations and the duration both change when a new operating point is selected. For the gates considered here, the lower peak Rydberg admixture of the longer pulses does not compensate for the decay accumulated during the longer evolution. Increasing $\alpha_0\tau_{\rm gate}$ beyond the value used in the main text therefore reduces $P_{\rm Ryd}^{\max}$ but increases the finite-lifetime error. The gate duration must consequently be chosen together with the allowed Rydberg population, rather than by minimizing the peak admixture alone.

%%%%%%%%%%%%%%%%%%%%%%%%%%%%%%%%%%%%%%%%%%%%%%%%%%%
\section{Spectator-shift model and crosstalk diagnostics}
\label{app:crosstalk}
%%%%%%%%%%%%%%%%%%%%%%%%%%%%%%%%%%%%%%%%%%%%%%%%%%%

In \sref{sec:crosstalk}, we calculate how the transient Rydberg excitation of a weakly driven spectator changes the conditional phase and fidelity of the addressed gate. Here, we give the three-atom Hamiltonian and explain how the phase and fidelity are obtained from the resulting evolution. We also examine their dependence on the leaked intensity and spectator interactions.

We use the dimensionless variables of \sref{sec:dimensionless}, with $\tau=\Delta t$, $\alpha(\tau)=\Omega(t)/\Delta$, and $\mathcal H=H/(\hbar\Delta)$. The time evolution is calculated in the three-atom bare-state basis specified below.

%%%%%%%%%%%%%%%%%%%%%%%%%%%%%%%%%%%%%%%%%%%%%%%%%%%
\subsection{Three-atom Hamiltonian}
%%%%%%%%%%%%%%%%%%%%%%%%%%%%%%%%%%%%%%%%%%%%%%%%%%%

We denote the gate atoms by $A$ and $B$ and the spectator by $C$. When $V_{AC}\neq V_{BC}$, the states with a Rydberg excitation on atom $A$ or atom $B$ acquire different spectator-induced shifts. The symmetric state $\ket{R}$ alone is then insufficient to describe the singly excited states. We therefore retain $\ket{1r_B}$ and $\ket{r_A1}$ separately and use the addressed-pair basis
\begin{equation}
\begin{split}
\mathcal B_{AB}=\{&
\ket{00},\ket{01},\ket{0r_B},
\ket{10},\ket{r_A0},\ket{11},\\
&\ket{1r_B},\ket{r_A1},\ket{rr},\ket{pp'}\}.
\end{split}
\label{eq:AB_basis_app}
\end{equation}
Together with the spectator basis $\mathcal B_C=\{\ket{1}_C,\ket{r}_C\}$, the three-atom basis contains 20 states.

The nonzero laser and F\"orster couplings in the addressed-pair Hamiltonian are
\begin{align}
\langle0r_B|\mathcal H_{AB}|01\rangle
&=
\langle r_A0|\mathcal H_{AB}|10\rangle
=
\frac{\alpha(\tau)}{2},
\nonumber\\
\langle1r_B|\mathcal H_{AB}|11\rangle
&=
\langle r_A1|\mathcal H_{AB}|11\rangle
=
\frac{\alpha(\tau)}{2},
\nonumber\\
\langle rr|\mathcal H_{AB}|1r_B\rangle
&=
\langle rr|\mathcal H_{AB}|r_A1\rangle
=
\frac{\alpha(\tau)}{2},
\nonumber\\
\langle pp'|\mathcal H_{AB}|rr\rangle
&=\beta,
\label{eq:HAB_offdiag_app}
\end{align}
with the Hermitian-conjugate matrix elements understood. The nonzero diagonal elements are
\begin{align}
\langle0r_B|\mathcal H_{AB}|0r_B\rangle
&=
\langle r_A0|\mathcal H_{AB}|r_A0\rangle
=
\langle1r_B|\mathcal H_{AB}|1r_B\rangle
\nonumber\\
&=
\langle r_A1|\mathcal H_{AB}|r_A1\rangle
=1,
\nonumber\\
\langle rr|\mathcal H_{AB}|rr\rangle&=2,
\qquad
\langle pp'|\mathcal H_{AB}|pp'\rangle=2+\gamma.
\label{eq:HAB_diag_app}
\end{align}

When the spectator interactions vanish, the equal laser couplings connect $\ket{11}$ to the symmetric combination of $\ket{1r_B}$ and $\ket{r_A1}$. The evolution from $\ket{11}$ then reduces to the four-level dynamics of \eref{eq:two_atom_hamiltonian}. Retaining the two singly excited states individually allows the same description to include unequal spectator shifts without imposing this symmetry.

The spectator is driven by the leaked field of \eref{eq:OmegaC}, with the same detuning as the gate atoms. In the basis $\{\ket{1}_C,\ket{r}_C\}$, its dimensionless Hamiltonian is
\begin{equation}
\mathcal H_C(\tau)
=
\begin{pmatrix}
0 & \alpha_C(\tau)/2\\
\alpha_C(\tau)/2 & 1
\end{pmatrix},
\qquad
\alpha_C(\tau)=\sqrt{\epsilon}\,\alpha(\tau).
\label{eq:HC_app}
\end{equation}
Combining the addressed-pair and spectator Hamiltonians gives the dimensionless form of \eref{eq:HABC},
\begin{equation}
\mathcal H_{ABC}(\tau)
=
\mathcal H_{AB}(\tau)\otimes I_C
+
I_{AB}\otimes\mathcal H_C(\tau)
+
\mathcal H_{\rm int}^{(C)}.
\label{eq:HABC_app}
\end{equation}
The spectator interaction contributes only when atom $C$ occupies $\ket{r}_C$, so we write
\begin{equation}
\mathcal H_{\rm int}^{(C)}
=
\mathcal H_{\rm shift}^{(AB)}
\otimes
\ket{r}_C\bra{r}_C,
\label{eq:HintC_app}
\end{equation}
where
\begin{align}
\mathcal H_{\rm shift}^{(AB)}
={}&
 v_0\left(\ket{0r_B}\bra{0r_B}+\ket{1r_B}\bra{1r_B}\right)
\nonumber\\
&+
\zeta v_0\left(\ket{r_A0}\bra{r_A0}+\ket{r_A1}\bra{r_A1}\right)
\nonumber\\
&+
(1+\zeta)v_0\ket{rr}\bra{rr}
+
\eta(1+\zeta)v_0\ket{pp'}\bra{pp'}.
\label{eq:Hshift_app}
\end{align}
The first two terms give the shifts of states containing a Rydberg excitation on atom $B$ and atom $A$, respectively. Both interactions contribute when the addressed pair occupies $\ket{rr}$, giving the sum $(1+\zeta)v_0$. The factor $\eta$ allows the spectator-induced shift of $\ket{pp'}$ to differ from this sum, as described in \sref{sec:crosstalk}. Because \eref{eq:HintC_app} contains the projector onto $\ket{r}_C$, all these shifts vanish when the spectator is in $\ket{1}_C$.

For isotropic van der Waals interactions with the same $C_6$ coefficient for the two spectator-gate pairs,
\begin{equation}
\zeta
=
\frac{V_{AC}}{V_{BC}}
=
\left(\frac{R_{BC}}{R_{AC}}\right)^6.
\label{eq:zeta_geometry_app}
\end{equation}
We vary $\zeta$ independently in the effective model to compare unequal and equal spectator interactions without selecting a particular three-atom geometry.

%%%%%%%%%%%%%%%%%%%%%%%%%%%%%%%%%%%%%%%%%%%%%%%%%%%
\subsection{Evaluation of phase and fidelity}
%%%%%%%%%%%%%%%%%%%%%%%%%%%%%%%%%%%%%%%%%%%%%%%%%%%

For each computational input $\ket{ij}$ of the addressed atoms, the three-atom system is initialized in $\ket{ij}\ket{1}_C$ and evolves under Eq.~\eqref{eq:HABC_app}. We denote the resulting three-atom state by $\ket{\Psi_{ij}(\tau)}$. The return amplitude at the end of the pulse is
\begin{equation}
a_{ij}
=
\langle ij,1_C|
\Psi_{ij}(\tau_{\rm gate})\rangle,
\qquad
\phi_{ij}
=
\arg(a_{ij}).
\label{eq:return_phase_app}
\end{equation}
Here, $\phi_{ij}$ is the phase acquired by the component that returns to the initial three-atom state $\ket{ij}\ket{1}_C$. The four return phases are combined as in Eq.~\eqref{eq:phi_cond_xtalk} to obtain the conditional phase of the addressed two-qubit gate in the presence of the spectator. The spectator-induced change in this phase is then calculated from Eq.~\eqref{eq:Deltaphi_xtalk} by subtracting the conditional phase of the spectator-free gate.

For the Bell-state fidelity, we propagate the initial state
$\ket{++}_{AB}\ket{1}_C$. After the pulse, the spectator atom is traced out,
\begin{equation}
\rho_{AB}
=
{\rm Tr}_C
\left[
\ket{\Psi_+(\tau_{\rm gate})}
\bra{\Psi_+(\tau_{\rm gate})}
\right].
\label{eq:rhoAB_app}
\end{equation}
We retain the $4\times4$ block of $\rho_{AB}$ corresponding to the
computational states $\{\ket{00},\ket{01},\ket{10},\ket{11}\}$ and denote it by $\rho_{\rm q}$. This block is not renormalized, so any population that does not return to the addressed-qubit computational subspace reduces the fidelity.

To compare with the spectator-free CZ gate, we apply the same
single-qubit phase correction used in Sec.~\ref{sec:pulse_metrics},
\begin{equation}
U_Z
=
Z(\theta_s^{(0)})\otimes Z(\theta_s^{(0)}),
\qquad
Z(\theta)
=
{\rm diag}(1,e^{-i\theta}),
\label{eq:UZ_xtalk_app}
\end{equation}
where $\theta_s^{(0)}$ is the single-atom return phase of the
spectator-free gate. The Bell-state fidelity in the presence of the spectator is then
\begin{equation}
F_{\rm Bell}^{({\rm xtalk})}
=
\langle\Phi_{\rm CZ}|
\rho'
|\Phi_{\rm CZ}\rangle,
\label{eq:Fbell_xtalk_app}
\end{equation}
where $\rho'=U_Z\rho_q U_Z^{\dagger}$.

We use the same local phase correction as for the spectator-free gate and do not readjust it for each spectator interaction. Consequently, spectator-induced changes in the single-qubit phases of atoms $A$ and $B$ contribute to $F_{\rm Bell}^{({\rm xtalk})}$, although these phases cancel from the conditional phase combination in \eref{eq:phi_cond_xtalk}. The fidelity therefore includes changes in both the single-qubit and conditional phases, as well as any population that does not return to the computational subspace of the addressed atoms.

Setting $v_0=0$ provides a direct check of the three-atom calculation. In this limit, the spectator is decoupled from the addressed atoms, so the total evolution factorizes even though the leaked laser field still drives atom $C$. After tracing over the spectator, the evolution of atoms $A$ and $B$ therefore reduces to the spectator-free two-atom result.

We also verify the spectator excitation independently using the
weak-dressing expression $P_{r_C}\simeq\epsilon\alpha^2/4$.
In \frefp{fig:crosstalk}{a}, we compare this expression with the
numerically calculated spectator population for the parameters used in the main text and find close agreement throughout the pulse.

\begin{figure}[t]
    \centering
    \includegraphics[width=\columnwidth]{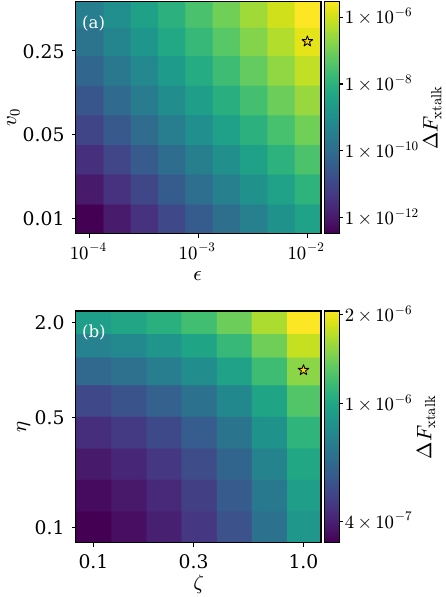}
    \caption{Dependence of the spectator-induced Bell-state fidelity loss
    $\Delta F_{\rm xtalk}$ on the leaked intensity and spectator
    interactions at the selected $\gamma=0$ gate operating point.
    (a) Variation with $\epsilon$ and $v_0=V_{BC}/\Delta$ at fixed
    $\zeta=1$ and $\eta=1$.
    (b) Variation with $\zeta$ and $\eta$ at fixed
    $\epsilon=10^{-2}$ and $v_0=0.3$.
    The stars mark the parameter set
    $\epsilon=10^{-2}$, $v_0=0.3$, $\zeta=1$, and $\eta=1$
    used in \sref{sec:crosstalk}.
    }
    \label{fig:breakdown}
\end{figure}
%

%%%%%%%%%%%%%%%%%%%%%%%%%%%%%%%%%%%%%%%%%%%%%%%%%%%
\subsection{Dependence on spectator parameters}
%%%%%%%%%%%%%%%%%%%%%%%%%%%%%%%%%%%%%%%%%%%%%%%%%%%

The calculation in \sref{sec:crosstalk} uses $\epsilon=10^{-2}$, $v_0=0.3$, $\zeta=1$, and $\eta=1$. To examine how the crosstalk depends on these choices, we vary the leaked intensity and spectator interactions at the selected $\gamma=0$ gate operating point. We consider $\epsilon$ and $v_0$ together while holding $\zeta=\eta=1$, and then vary $\zeta$ and $\eta$ with $\epsilon=10^{-2}$ and $v_0=0.3$. The gate pulse and addressed-pair parameters are unchanged in these calculations. This allows the effect of the spectator parameters to be compared for the same two-qubit gate.

In \frefp{fig:breakdown}{a}, we show the dependence of $\Delta F_{\rm xtalk}$ on the leaked intensity and the spectator interaction. The fidelity loss increases with both $\epsilon$ and $v_0$. A larger leaked intensity increases the transient Rydberg population of atom $C$, whereas a larger $v_0$ increases the interaction shifts when the spectator is excited. Both effects increase the additional phases acquired by the addressed atoms during the pulse. For weak leakage and small spectator interactions, the leading conditional phase change is proportional to $\epsilon v_0$. When the unperturbed gate is close to the ideal CZ operation and the error is dominated by these small phase changes, the leading fidelity loss is quadratic in the phase. This accounts for the increase of the crosstalk loss with both parameters. The numerical results remain below $3.0\times10^{-6}$ over the range examined.

In \frefp{fig:breakdown}{b}, the dependence on $\zeta$ and $\eta$ is shown. Changing $\zeta$ changes the spectator interaction with atom $A$ relative to that with atom $B$, while changing $\eta$ changes the spectator-induced shift of $\ket{pp'}$ relative to that of $\ket{rr}$. These parameters therefore affect the phases through different terms of \eref{eq:Hshift_app}. Over the range shown, the fidelity loss increases with both parameters and lies between $3.4\times10^{-7}$ and $2.1\times10^{-6}$. The small crosstalk contribution is thus not restricted to equal spectator interactions or to equal shifts of the two Rydberg pair states.

For comparison, the finite-lifetime contribution at $\Gamma/\Delta=5\times10^{-5}$ is approximately $1.5\times10^{-3}$, as discussed in \sref{sec:decay}. The spectator-induced fidelity loss remains well below this value throughout both parameter ranges in \fref{fig:breakdown}. These calculations support the comparison made in \sref{sec:crosstalk} for weak illumination and the spectator interactions considered here. The conclusion applies to the diagonal shift model and the ranges examined, without requiring a particular choice of $\zeta$ or $\eta$ within those ranges.
\bibliographystyle{apsrev4-1}
\bibliography{references}
%%%%%%%%%%%%%%%%%%%%%%%%%%%%%%%%%%%%%%%%%%%%%%%%%%%
\end{document}